\documentclass[10pt]{article}
\usepackage{epsfig}

\newcommand{\ud}{\mathrm{d}}

\begin{document}

    \title{\LARGE \bf  Self-Dual Cosmic Strings and Gravitating Vortices in
    Gauged Sigma Models \\}

   \author{
  \large  Y. Verbin$^a$\thanks{Electronic address:
verbin@oumail.openu.ac.il}$\:$,
  S. Madsen$^b$\thanks{Electronic address:  Soeren@Madsen2TheMax.dk} $\:$,
 A.L. Larsen$^b$\thanks{Electronic address: all@fysik.sdu.dk}$\:$}
 \date{ }
   \maketitle
 \centerline{$^a$ \em Department of Natural Sciences, The Open University
   of Israel,}
   \centerline{\em P.O.B. 39328, Tel Aviv 61392, Israel}
     \vskip 0.4cm
   \centerline{$^b$ \em Department of Physics, University of Odense, }
   \centerline{\em Campusvej 55, 5230 Odense M, Denmark}
   \vskip 1.1cm

   \begin{abstract}
Cosmic strings are considered in two types of gauged sigma models,
which generalize the gravitating Abelian Higgs model. The two
models differ by whether the U(1) kinetic term is of the Maxwell
or Chern-Simons form. We obtain the self-duality conditions for a
general two-dimensional target space defined in terms of field
dependent "dielectric functions". In particular, we analyze
analytically and numerically the equations for the case of O(3)
models (two-sphere as target space), and find cosmic string
solutions of several kinds as well as gravitating vortices. We
classify the solutions by their flux and topological charge. We
note an interesting connection between the Maxwell and
Chern-Simons type models, which is responsible for simple
relations between the self-dual solutions of both types. There is
however a significant difference between the two systems, in that
only the Chern-Simons type sigma model gives rise to
spinning cosmic vortices.\\
   \end{abstract}

   {\em PACS numbers: 11.27.+d, 04.20.Jb, 11.10.Kk,  11.10.Lm. }

\section{Introduction}

Topological defects are generally believed to have been formed
during a series of phase transitions in the early universe. In
particular, many field-theoretical models suggest the formation of
cosmic strings (for a review, see for instance Vilenkin and
Shellard \cite{VilSh}).

The BOOMERANG results \cite{2}, especially concerning the location
of the acoustic peaks \cite{3,4}, essentially rule out GUT scale
($\sim10^{16}$GeV) topological defects being the {\it sole} source
of the cosmic microwave background (CMB) anisotropy, which
reflects the early universe density fluctuations, eventually
leading to galaxy formation. However, recent analysis
\cite{5,6,7,8,9} shows that a mixture of cosmic strings and
inflation is consistent with current CMB data.

Independently of a possible role in the large scale structure
formation, and even if their mass scale is lower than the GUT
scale, cosmic strings may have also a number of significant
observable astrophysical effects. Just to mention a few: Cosmic
strings could be sources of double images \cite{VilSh}, of
gravitational waves \cite{VilenkinGW,DamourVilenkin} and of ultra
high energy cosmic rays \cite{BonazzolaPeter}. Recently it was
even suggested \cite{BerezinskyEtAl} that they can serve as gamma
ray burst engines. Thus, cosmic strings are still of considerable
interest in cosmology and astrophysics.

The most popular field-theoretical system, used to model cosmic
string generation \cite{VilSh}, is the Abelian Higgs model. There
exist, however, other systems which may be used for the same
purpose and are not more complicated. As long as we model cosmic
strings as static cylindrically symmetric sources coupled to
gravity, which is essentially a two dimensional system, we may
borrow circular solutions of other well-known two dimensional
systems, namely non-linear sigma models.

Non-linear sigma models \cite{Rajaraman} are very much used as
effective theories describing various systems such as low energy
effective QCD, and are also used as "toy models" for gauge
theories. A generalization of these, where the global symmetry
group or its subgroup is promoted to be local, are gauged sigma
models which appear naturally in supersymmetric field theories.

All these models have been extensively studied for decades. Most
of the results were obtained in flat (Minkowski) spacetime,  but
more recently gravitating solutions were studied too. However, the
gravitating sigma model solutions are usually taken to be
spherically symmetric as in the cases of gravitating Skyrmions and
textures (see e.g. Volkov and Gal'tsov \cite{Volkov}), although some
authors gave attention to cylindrically symmetric solutions of the
non-gauged \cite{ComtetGibbons,KimMoon} as well as the gauged
models with either Maxwell \cite{YYang} or Chern-Simons
terms \cite{Valtancoli,London,Clement,Chung2Kim2}.

Here we will analyze the gravitating gauged sigma models with a
general two-dimensional target space defined in terms of two field
dependent "dielectric functions", which may be viewed as
generalized Abelian Higgs models: Maxwell type and  Chern-Simons
type, according to the gauge field term in the Lagrangian. We will
find the general conditions for self-dual cosmic string solutions
and get several kinds of interesting solutions.

We will start by considering, in the next section, static and
translationally invariant solutions of a generic gauged
gravitating non-linear sigma model of a Maxwell type, and find
simple conditions for the existence of self-dual solutions. In
sections 3 and 4, we concentrate in the O(3) model and add
rotational symmetry to obtain self-dual cosmic string solutions.
In section 5, we move to the Chern-Simons type of model, and
consider stationary (not static) solutions. First we obtain the
self-duality conditions for this kind of solutions, and then we
specialize to spinning cosmic vortices. In section 6, we discuss
the relation between the two types of theories and solutions, and
we end with a discussion of the topological charges in section 7.

\section{The Generalized Abelian Higgs Model}
\setcounter{equation}{0}

We start by considering a generalization of the Abelian Higgs
model defined by the action:
\begin{eqnarray}
S=\int \ud^4x\sqrt{|g|}\left( \frac{1}{2}{\cal E}_1(|\Phi|)(D_\mu
\Phi)^*(D^\mu
\Phi)-U(|\Phi|)-\frac{1}{4}{\cal E}_2(|\Phi|)F_{\mu\nu}F^{\mu\nu}+\frac{1}{16\pi
G}R\right) \label{action1}
\end{eqnarray}
where ${\cal E}_1(|\Phi|)$ and ${\cal E}_2(|\Phi|)$ are
non-negative dimensionless functions, which may be interpreted as
Weyl factors of a conformally-flat target space metric of a
non-linear sigma model with local U(1) symmetry \cite{susy}. The
function ${\cal E}_2(|\Phi|)$ plays further the role of a dielectric
function \cite{LeeNam,GibbonsWells}. In what follows, we use the
collective notation ${\cal E}_a,\; (a=1,2$), and refer to both as
"dielectric functions".
  The action (\ref{action1}) also generalizes the
ones considered by Lohe \cite{lohe1, lohe2}. Furthermore, the action
(\ref{action1}) arises in various extended supergravity theories
\cite{susy}. In the general case, we may also add the term ${\cal
E}_3(|\Phi|)\epsilon^{\mu\nu\rho\sigma}F_{\mu\nu}F_{\rho\sigma}$,
but this term vanishes identically for the configurations
considered in the present paper and has no effect on the field
equations.

The field equations derived from the action (\ref{action1}) are:

\begin{equation}
{\cal E}_1 (|\Phi|) D_\mu D^\mu \Phi + \frac{\Phi^*}{2|\Phi|}
\frac{d{\cal E}_1}{d|\Phi|} D_\mu \Phi D^\mu \Phi +
\frac{\Phi}{|\Phi|} \frac{dU}{d|\Phi|} + \frac{\Phi}{4|\Phi|}
\frac{d{\cal E}_2}{d|\Phi|}F_{\mu\nu}F^{\mu\nu}= 0 \label{NLHiggs}
\end{equation}
\begin{equation}
\nabla_\mu \left({\cal E}_2(|\Phi|) F^{\mu\nu} \right)=j^{\nu} =
-\frac {i}{2}e{{\cal E}_1} (|\Phi|)(\Phi^*(D^{\nu}\Phi) - \Phi
(D^{\nu}\Phi)^*) \label{NLMax}
\end{equation}
\begin{eqnarray}
\frac{1}{8\pi G}R_{\mu\nu}+\frac{1}{2}{\cal E}_1 (|\Phi|) \left(
(D_\mu \Phi)^*(D_\nu\Phi)+(D_\nu \Phi)^*(D_\mu\Phi)
\right)-U(|\Phi|)g_{\mu\nu} + \nonumber \\ {\cal
E}_2(|\Phi|)\left(g^{\kappa\lambda} F_{\mu\kappa}F_{\lambda\nu} +
\frac{1}{4}F_{\kappa\lambda}F^{\kappa\lambda}g_{\mu\nu} \right)= 0
\label{EinstNLH}
\end{eqnarray}
Conventions: $D_\mu = \nabla_\mu - ieA_\mu$, signature $(+,-,-,-)$
and $R^\kappa_{ \lambda\mu\nu} =
\partial_\nu\Gamma^\kappa_{\lambda\mu} -
\partial_\mu\Gamma^\kappa_{\lambda\nu}+ ..$.

Usually (but not always) we will be interested in potentials,
which ensure spontaneous symmetry breaking and which lead to a
massive gauge field. That is to say, potentials with a circle of
degenerate minima at (say) $| \Phi |=v$ and with positive second
derivative at $v$. We further normalize the potential such that it
vanishes in the vacuum. Thus, altogether:
\begin{equation}
U=0,  \ \ \ \ U'=\frac{dU}{d|\Phi|}=0;   \
 \;\; \ \mbox{for}\;\; \ \ |\Phi|=v
\end{equation}
With these normalizations, the Higgs mass and gauge boson mass are,
respectively:
\begin{equation}
m_H^2=\frac{U''(v)}{{\cal E}_1(v)}, \;\;\;\;
m_A^2=\frac{e^2v^2{\cal E}_1(v)}{{\cal E}_2(v)}\label{SSBMasses}
\end{equation}

In order to study cosmic string solutions, we assume the metric and
matter fields to be static and symmetric under $z$-translations.
Thus we assume that $A_\mu$ and $\Phi$ depend only on the two
transverse coordinates $x{^k}$, and for the metric we use the
following form:

\begin{equation}
ds^2=N^2(x{^k})dt^2- \gamma_{ij}
(x{^k})dx{^i}dx{^j}-K^2(x{^k})dz^2 \label{staticMetric}
\end{equation}
We also require the presence of a magnetic field only, i.e.  that
the gauge potential will have the form:
\begin{equation}
A_{\mu}dx^\mu = A_i(x{^k}) dx^i \label{GaugeField4D}
\end{equation}
such that the Maxwell tensor will contain a single magnetic
component $B$:
\begin{equation}
F_{\mu \nu}dx^\mu\wedge dx^\nu = - B {\sqrt{|\gamma|}}
\epsilon_{ij}dx^i\wedge dx^j \label{Maxwell2Form}
\end{equation}
where $|\gamma|=|{\mbox{det}}(\gamma_{ij})|$. In order to get the
field equations for static solutions, we compute also the
components of the Ricci tensor:
\begin{eqnarray}
R_{00}&=&-\frac {N}{K}\nabla_i(K \nabla^{i} N) \nonumber\ \\
R_{33}&=&\frac {K}{N}\nabla_i(N \nabla^{i} K) \nonumber\ \\
R_{ij}&=&R_{ij}(\gamma)+\frac{1}{N}\nabla_i \nabla_{j} N +
\frac{1}{K}\nabla_i \nabla_{j} K \nonumber\ \\
R_{i0}&=&R_{i3}= 0, \;\;\;\; R_{03}=0 \label{Ricci4Dstatic}
\end{eqnarray}
where $\nabla_i$ is the covariant derivative with respect to the
two-dimensional metric $\gamma_{ij}$ and $R_{ij}(\gamma)$ the
corresponding Ricci tensor.

A significant simplification of this system is obtained if
self-duality conditions are satisfied, i.e. if the system admits a
Bogomolnyi limit \cite{Bogomolnyi}. It is well-known
\cite{ComtetGibbons} that in the usual Higgs model, the
flat space considerations can be carried over to curved background
if $N(x^i)$ and $K(x^i)$ are constants, say, 1. We will see now
that the present generalized Higgs model has also a Bogomolnyi
limit if we keep $N(x^i)=K(x^i)=1$. If we use these conditions, we
find that the $(00)$ and $(33)$ components of Einstein equations
will be satisfied only if:
\begin{equation}
U(|\Phi|)= \frac{1}{2}{\cal E}_2(|\Phi|) B^2
 \label{equB}
 \end{equation}
Now we turn to the $(ij)$ components of the Einstein equations, or
even better to $G_{ij}$ which vanish identically. Consequently,
$T_{ij}=0$ as well and we get:
\begin{eqnarray}
\frac{1}{2}{\cal E}_1(|\Phi|)\left( (D_i \Phi)^*(D_j\Phi)+(D_j
\Phi)^*(D_i\Phi) \right)- \nonumber\ \\
\left( \frac{1}{2}{\cal E}_1
(|\Phi|) \gamma ^{kl}(D_k \Phi)^*(D_l\Phi)+U(|\Phi|)
 -\frac{1}{2}{\cal
E}_2(|\Phi|) B^2 \right)\gamma_{ij} = 0 \label{Einst4ij}
\end{eqnarray}
Using (\ref{equB}), this condition simplifies further and it
follows that it is equivalent to the curved spacetime version of
the self-duality condition:
\begin{equation}
D_i\Phi=i\eta {\sqrt{|\gamma|}} \epsilon_{ij} \gamma^{jk} D_k\Phi
 \label{selfduality}
 \end{equation}
 where  $\eta=\pm 1$ corresponds to self-dual or anti
 self-dual solutions. This is a first order equation for the Higgs field.
  Analogously, Eq.(\ref{equB}) is a first order equation for the gauge
potential.

We also find the following expression for the two dimensional
Ricci scalar:
\begin{equation}
R(\gamma)= -8\pi G \left({\cal E}_1 (|\Phi|) \gamma^{ij}(D_i
\Phi)^*(D_j\Phi) + 4U(|\Phi|)\right)
 \label{RicciNLHSD}
 \end{equation}
which serves as the Einstein equation for the two metric
$\gamma_{ij}$.

Equation (\ref{NLHiggs}) for the Higgs field becomes a consistency
condition, which constrains the form of the potential $U(|\Phi|)$:
\begin{equation}
\eta e|\Phi| {\cal E}_1(|\Phi|) B +\frac{dU}{d|\Phi|}+
\frac{B^2}{2}\frac{d{\cal E}_2}{d|\Phi|}=0
 \label{Higgs2}
\end{equation}
  Maxwell equations (\ref{NLMax}) give:
 \begin{equation}
\partial_j ({\cal E}_2(|\Phi|) B) = -\frac
{\eta e}{2}{\cal E}_1(|\Phi|) \partial_j |\Phi|^2 \label{Maxwell2}
\end{equation}
which is a second order equation for the gauge potential, but it is
not an independent one, as is easily shown: If we think of $B$ as a
function of $|\Phi|$, we may use (\ref{equB}) and get from
(\ref{Higgs2}) the following relation, which we will refer to as
the "Bogomolnyi constraint":
\begin{equation}
\frac{d}{d|\Phi|} ({\cal E}_2(|\Phi|) B) +\eta e |\Phi| {\cal E}_1(|\Phi|)=0
\label{Bogconstraint}
\end{equation}
This is actually again the Maxwell equation (\ref{Maxwell2}) in
disguise, and can be used together with
(\ref{equB}) to get the function $U(|\Phi|)$ for any set of
given "dielectric functions" ${\cal E}_a,\;\; a=1,2$. In order to do it
directly, we need to express the magnetic field $B$ in terms of the
potential $U$. If $B$ has a definite sign, we infer from
the Bogomolnyi constraint that ${\cal E}_2 (|\Phi|) B(|\Phi|)\geq 0$ for
 $\eta = +1$, and similarly for negative values, so we may take a square root
of (\ref{equB}):
\begin{equation}
 B =\eta \sqrt { 2U/ {\cal E}_2}
 \label{equBsqrt}
 \end{equation}
and get a simple relation between the potential and the two
dielectric functions, which is a necessary condition for the
action (\ref{action1}) to have self-dual solutions, i.e. a
Bogomolnyi limit:
\begin{equation}
\frac{d}{d|\Phi|}\sqrt {2{\cal E}_2(|\Phi|) U(|\Phi|)} + e |\Phi| {\cal E}_1(|\Phi|)=0
\label{BogconstraintForU}
\end{equation}

 The simplest case where we can apply (\ref{BogconstraintForU})
is of course ${\cal E}_a = 1$, which reproduces immediately the standard
Higgs potential:
\begin{equation}
 U(|\Phi|) = \frac{a}{4}(v^2-|\Phi|^2)^2
 \label{HiggsSelfdualPot}
\end{equation}
where the Bogomolnyi relation between the coupling constants
holds:
\begin{equation}
 \alpha\equiv e^2/a = 2
 \label{BogHiggs}
\end{equation}
Note that here, the vacuum expectation value $v$ enters into the
potential as an integration constant, and has also a role of the
value of $|\Phi|$ for which the magnetic field vanishes.

The next case is the non-linear O(3) sigma model, which is obtained
by taking ${\cal E}_1$ to be the usual conformal factor for
$\mbox{\textrm{S}}^2$ with a radius $\mu/2$ \cite{Rajaraman}:
\begin{equation}
 {\cal E}_1(|\Phi|) = 1/(1+|\Phi|^2/\mu^2)^2
 \label{so3conformalfactor}
\end{equation}
The parameter $\mu$ sets a second energy scale in the system and
renders ${\cal E}_1$ dimensionless. It is very simple to integrate
the Bogomolnyi constraint (\ref{Bogconstraint}), or equivalently
Eq.(\ref{BogconstraintForU}), also in this case and get the
following form of $B(|\Phi|)$ and potential, which allows a
Bogomolnyi limit in the O(3) generalized Abelian Higgs model:
\begin{equation}
B(|\Phi|) = \frac{\eta e}{2(1+v^2/\mu^2)}\frac{1}{{\cal
E}_2(|\Phi|)} \frac{v^2-|\Phi|^2}{1+|\Phi|^2/\mu^2}
 \label{SigmaSelfdualB}
\end{equation}

\begin{equation}
 U(|\Phi|) = \frac{e^2}{8(1+v^2/\mu^2)^2}\frac{1}{{\cal E}_2(|\Phi|)}
\frac{(v^2-|\Phi|^2)^2}{(1+|\Phi|^2/\mu^2)^2}
 \label{SigmaSelfdualPot}
\end{equation}
Here again $v$ is an integration constant, which parametrizes the
minimum of the potential and the field value for which the
magnetic field vanishes. Generally there is no relation between
this scale and the scale $\mu$.
We therefore have a one-parameter family of potentials in this
O(3) generalized Abelian Higgs model for any given ${\cal
E}_2(|\Phi|)$, which we may take at our will. Some special cases
with ${\cal E}_2(|\Phi|)=1$ were already analyzed by several
authors mainly in flat space \cite{Schr1,Mukh1}, but also with
coupling to gravity \cite{YYang}.

A more "symmetric" picture may be obtained by using the
angular variable on $\mbox{\textrm{S}}^2$ defined by:
\begin{equation}
|\Phi| = \mu \tan (\Theta/2) \label{stereographic}
\end{equation}
Generically, the potential exhibits a U(1)-symmetry breaking
minimum at $|\Phi|=v$, but there are two special limits $v=0,
\infty$, where the ground state is only a point in target space
(the north or south poles of $\mbox{\textrm{S}}^2$); namely, no
symmetry breaking occurs. This pattern of symmetry breaking is
reflected by the Higgs and gauge boson masses, which turn out to
have the following equal (due to the self-duality) values:
\begin{equation}
 m_H=m_A= \frac{ev}{1+v^2/\mu^2}
\label{HiggsGaugeMass}
\end{equation}
The two cases of symmetric vacuum correspond to vanishing Higgs
and gauge masses.

Just as a final check we note that the Higgs potential
((\ref{HiggsSelfdualPot}) with (\ref{BogHiggs})) is obtained in
the limit $\mu\rightarrow \infty$ from (\ref{SigmaSelfdualPot})
with ${\cal E}_2(|\Phi|)=1$. The masses which are obtained in this limit
are the usual ones.

Now we return to the Einstein equation (\ref{RicciNLHSD}) and
notice that for self-dual solutions, its right hand side is
actually a two-dimensional divergence (it must be as we will see
in section \ref{Topological Charge}), and may be simplified to:
\begin{equation}
R(\gamma)= \frac{16\pi G}{\eta e} \nabla_{i}\left(\gamma^{ij}{\cal
E}_2(|\Phi|) B
\partial_j \log|\Phi|\right)
\label{SimplifRicciNLHSD}
\end{equation}
Without loss of generality, we may take the two-dimensional
transverse part of the metric tensor to be conformally flat,  i.e.
$\gamma_{ij}=H^2(x^k)\delta_{ij}$ so the Einstein equation
(\ref{RicciNLHSD}) reduces further to:
\begin{equation}
\delta^{ij}\partial_i \left( \partial_j\log H -\frac{8\pi G}{\eta
e}{\cal E}_2(|\Phi|) B \partial_j \log|\Phi|\right)=0
\label{Einstein2}
\end{equation}
This equation is not of first order, but it may get the form of
 a two dimensional Laplace equation, if we introduce a
 "super dielectric function" ${\cal A}(|\Phi|)$, which solves the following
equation:
\begin{equation}
|\Phi|\frac{d{\cal A}}{d|\Phi|} = \eta {\cal E}_2(|\Phi|)
B(|\Phi|)= \sqrt {2{\cal E}_2(|\Phi|) U(|\Phi|)}
\label{superdielectric1st}
\end{equation}
or equivalently the second order equation:
\begin{equation}
\frac{1}{|\Phi|} \frac{d}{d|\Phi|}\left( |\Phi|\frac{d{\cal A}}{d|\Phi|} \right) +
 e {\cal E}_1(|\Phi|)=0
\label{superdielectric2nd}
\end{equation}
Now we can use the function  ${\cal A}$, in order to give Eq.
(\ref{Einstein2}) the form of a Laplace equation:
\begin{equation}
\delta^{ij}\partial_i \partial_j \left(\log H -\frac{8\pi
G}{e} {\cal A} \right)=0
\label{EinsteinLaplace}
\end{equation}
Special solutions will be discussed in section 4.

\section{Flat Space Solutions}
\setcounter{equation}{0}
\label{Flat Space Solutions}

 First we discuss flat space solutions within this general
framework.  As mentioned above, some of the solutions are already
known but new ones can be easily obtained.

In order to study a single cosmic string solution, we take the usual
cylindrically symmetric Nielsen-Olesen ansatz for $n$ flux
units:
\begin{equation}
\Phi=\mu f(r)e^{in\varphi}\  ,\;\;\; A_\mu dx^\mu=A(r)d\varphi
\label{cylAnsatz}
\end{equation}
 where $\mu$ is a second energy scale, which is generally independent
 of $v$.  We will also assume that the dielectric functions depend on
$|\Phi|$ through the dimensionless ratio  $|\Phi|/\mu = f$ only.
For further use we define here $v/\mu\equiv\beta$.

As for the boundary conditions, we will see that they will not be
the same for all systems, but will rather have to be adapted to the
specific system. However, since we are interested in solutions
with finite energy per unit length and finite flux, the boundary
conditions at infinity should ensure asymptotically vanishing
energy density. The usual Nielsen-Olesen conditions should be generalized
such that all three contributions (scalar gauged kinetic term, potential and
Maxwell term) will vanish asymptotically. That is:
\begin{equation}
\lim_{r\rightarrow \infty} U(f(r))=0 ,\;\;\; \lim_{r\rightarrow
\infty} {\cal E}_2(f(r)) B^2 (r)=0 ,\;\;\; \lim_{r\rightarrow
\infty} f^2 (r) {\cal E}_1(f(r)) (eA(r)-n)^2=0 \label{BCatInf}
\end{equation}
In order to proceed, we will concentrate in the O(3) model. The
field equations (with $\eta=+1$ which we will mostly use from now
on) are the following simple first order set:
\begin{equation}
\frac{rf'}{f}=eA(r)-n \label{cylindrselfduality}
\end{equation}
\begin{equation}
\frac{A'}{r}=-\frac{e\mu^2}{2(1+\beta^2)}\frac{\beta^2-f^2}{{\cal
E}_2(f)(1+f^2)} \label{cylindrMax}
\end{equation}
where a prime denotes differentiation with respect to $r$.

We still have a freedom in the function ${\cal E}_2$, but
we choose until further notice ${\cal E}_2(f)=1$. The simplest
case, which yields the closest to the Abelian Higgs flux tube, is
the case $\beta=1$ which corresponds to a potential with a minimum
along the equator of $\mbox{\textrm{S}}^2$. We thus impose the
following additional boundary conditions:
\begin{equation}
f(0)=0\  ,\;\;\; A(0)=0 \label{moreBC}
\end{equation}
and the two last conditions at (\ref{BCatInf}) may be replaced by:
\begin{equation}
\lim_{r\rightarrow \infty} A(r)=n/e \label{BConAatInf}
\end{equation}
Note that the boundary conditions we took enforce $n$ to be
negative and $A(r)$ to be a non-positive decreasing function. The
magnetic field $B(r)$ is non-negative and the flux (in units of $2\pi/e$)
 is $-eA(\infty)=-n$, which is positive.

\begin{figure}[!t]
   \begin{center}
      \includegraphics[width=10cm,angle=270]{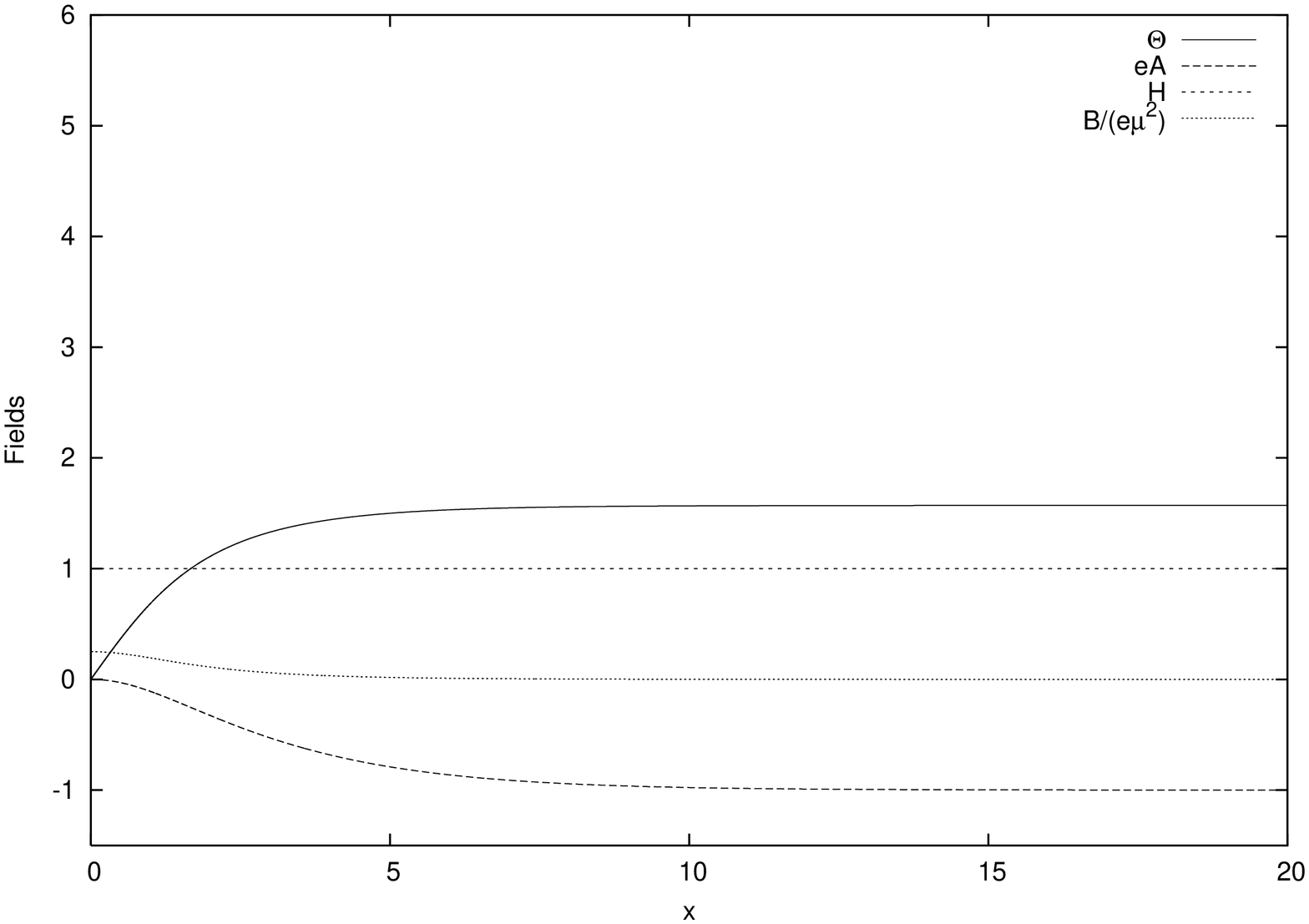} \\
   \caption{The solution to Eqs.(\ref{cylindrselfduality}),(\ref{cylindrMax}),
for ${\cal E}_2=1$. $\Theta$ is the angular coordinate on $S^2$.
We also show the dimensionless magnetic field. The metric field,
which is constant in this case, is included for comparison with
the following figures. The parameters used are  $n=-1,\ \beta =1$.
The dimensionless length coordinate is defined by $x=e\mu r$. }
 \label{figure1}
     \end{center}
     \end{figure}

There is no known analytical solution of this system, but it is
very easy to get numerical solutions. We will comment a little more
about it at the end of this section.
Fig.\ref{figure1} contains the field variables
($\Theta$, $A$ and $B$)  in this $\beta=1$ case.
This is
actually the solution discussed already by Mukherjee \cite{Mukh1}.
There is another related solution, which is obtained by reflecting
this solution in the  $\mbox{\textrm{S}}^2$ target space with
respect to the equator, i.e. starting at $f=\infty$ on the $z$-axis
($r=0$) and decreasing with $r$ to $f=1$:
\begin{equation}
\lim_{r\rightarrow 0} f(r)=\infty  ,\;\;\; \lim_{r\rightarrow
\infty} f(r)= 1\label{BCinvMukherjeeOnf}
\end{equation}
The reflection property is much more transparent in terms of the
angular field variable $\Theta$, Eq.(\ref{stereographic}), in
terms of which the boundary conditions on the scalar field are:
\begin{equation}
\Theta(0)=\pi  ,\;\;\; \lim_{r\rightarrow \infty} \Theta(r)=\pi/2
\label{BCinvMukherjeeonTheta}
\end{equation}
This reflected solution has obviously negative magnetic flux.

Another solution, which was discussed already by
Schroers \cite{Schr1}, corresponds to the limiting case of potential
with a minimum which does not break the U(1) symmetry. This
corresponds, in our terminology, to $\beta=0$ while $\mu$ stays
finite so $v=0$. In this case, the minimum of the potential is on
the north pole of $\mbox{\textrm{S}}^2$. Finite energy (per unit
length) solutions still exist, but they have now different boundary
conditions namely:
\begin{equation}
\Theta(0)=\pi  ,\;\;\; A(0)=0 \label{SchroersBCatOrigin}
\end{equation}
\begin{equation}
\lim_{r\rightarrow \infty} \Theta (r)=0\  ,\;\;\;
\lim_{r\rightarrow \infty} B(r)=0 \label{SchroersBCatInf}
\end{equation}
The last boundary condition in (\ref{BCatInf}) is automatically
 satisfied as  $f\rightarrow 0$ asymptotically.
This gives a solution, which maps the $z$-axis ($r=0$) to the
south pole of $\mbox{\textrm{S}}^2$ and the circle $r\rightarrow
\infty $ to the north pole. Since there is no symmetry breaking at
spatial infinity, the magnetic flux is not quantized but rather
takes continuous values.

In order to compare these solutions with the $\beta = 1$
ones, it is more instructive to present the "anti-Schroers"
solutions, which are solutions to the same problem with the
reflected potential ($\beta \rightarrow \infty$). The boundary
 conditions should be therefore also reflected:
\begin{equation}
\Theta(0)=0  ,\;\;\; A(0)=0 \label{AntiSchroersBCatOrigin}
\end{equation}
\begin{equation}
\lim_{r\rightarrow \infty} \Theta (r)=\pi\  ,\;\;\;
\lim_{r\rightarrow \infty} B(r)=0 \label{AntiSchroersBCatInf}
\end{equation}
where again the last boundary condition in (\ref{BCatInf}) is automatically
 satisfied, as now  $f^2 {\cal E}_1 \rightarrow 0$ asymptotically.
There are two unusual features of this family of solutions: First,
as mentioned above, there is no flux quantization, and second,
there does not exist a solution with $|n|=1$ but with $|n| \geq 2$
\cite{Schr1}. Consequently, the energy density is maximal not on
the symmetry axis, but rather on a cylindrical surface whose
radius is of the order of $1/e\mu$. Similar behavior is seen in
the analogous
 $|n| \geq 2$ Nielsen-Olesen flux tubes \cite{VilSh}.

Another kind of interesting potential is a triple well potential,
whose minima are at the north and south pole and at the equator of
the target space. In term of the angular field $\Theta$, it has the
following simple form:
\begin{equation}
U(\Theta) = \frac{e^2 \mu^4}{128}\sin ^2 (2\Theta)
\label{SigmaTripleWell}
\end{equation}
This potential is obtained by using ${\cal
E}_2(|\Phi|)=\frac{1}{4}(|\Phi|/\mu+\mu/|\Phi|)^2$ and $v=\mu$ in
Eq.(\ref{SigmaSelfdualPot}). Actually, ${\cal E}_2$ may be
multiplied by an arbitrary constant, thus having the height of the
potential free. Since the "polar" minima (at $\Theta =0,\pi$) do
not break the U(1) symmetry, we do not expect flux quantization
for solutions that approach asymptotically one of these minima. On
the other hand, solutions which have $\Theta = \pi/2$ as their
asymptotic value, will have of course quantized magnetic flux.

There are thus several kinds of finite energy (per unit length)
solutions in this system. For $\Theta (0) =0$, all three kinds
with asymptotic values $\Theta =0, \pi/2, \pi$ are realized. The
fields $\Theta$, $\bar{A}$ and $\bar{B}$ in figures
\ref{figure3},\ref{figure5},\ref{figure7} which actually
correspond to the Chern-Simons type system analyzed in sec.
\ref{GCSH Model} depict also the solutions of the present system
with $\Theta(\infty)=\pi/2, \pi, 0$ respectively. We will say more
about it in sec. \ref{correspondence}.

Similar solutions exist with $\Theta (0) =\pi$. A different kind
of solution starts with a "non-polar" value (i.e. $\Theta (0) \neq
0,\pi$), but for continuity must have $n=0$. The asymptotic values
correspond to each of the three possible vacua.

The numerical procedure we have used is essentially the same as in
previous works \cite{CLV,VMLC}, and we will not elaborate about it
more than the following remark regarding the treatment of boundary
conditions. A common practice in numerical analysis is to replace
the boundary conditions at infinity with the same boundary
conditions at some finite point, $r_\infty$. This works well on
the system (\ref{cylindrselfduality})-(\ref{cylindrMax}), provided
that $\beta$ is not $0$ or $\infty$. If we try to shoot from
$r_\infty$ to $0$ in these two cases, it is easy to see (from the
equations in terms of the angular field, $\Theta$), that the
solutions become constant. We are therefore forced to replace the
boundary condition on $\Theta$ with
$\Theta(\infty)=\beta\pm\delta$ where $\delta$ is a (positive)
parameter, which is to be determined numerically. It turns out,
that the continuum of solutions is found by giving $\delta$
slightly different values.

Another remark concerns the classification of solutions according
to the possible choices of boundary conditions. As is already
clear from the flat space solutions, the O(3) system has a
reflection symmetry with respect to the $\mbox{\textrm{S}}^2$
equatorial plane (i.e. $\Theta\rightarrow \pi-\Theta$ or
$f\rightarrow 1/f$) together with $\beta\rightarrow 1/\beta$. We
may therefore limit ourselves to solutions with $f(0)=0$ (which
for $\eta=+1$ are also $n<0$ solutions) and classify them
according to the asymptotic values $f(\infty)$, which are the
various minima of the potential function. The $n>0$ solutions for
the same $\eta$ and $\beta$ (and hence the same potential), which
start at the "south pole" of target space, are easily obtained
from the $n<0$ solutions with $\beta \rightarrow 1/\beta$ and
further obvious changes as $n\rightarrow -n$ and $A\rightarrow
-A$.\vspace{5mm}

\section{Gravitating Solutions}
\setcounter{equation}{0}
 In this section we consider gravitating O(3) cosmic strings.
 It is in fact very easy to get the gravitating self-dual
cylindrical solutions (i.e. single cosmic string solutions) in the
Bogomolnyi limit, since the field equations for the matter (scalar
and gauge) fields are very similar to those in flat spacetime.
More precisely, they are given by the following first order
equations:
\begin{equation}
\frac{rf'}{f}=\eta \left(eA(r)-n\right)
\label{Gravcylindrselfduality}
\end{equation}
which is identical with the Minkowskian one (for $\eta =+1$
compare (\ref{cylindrselfduality})), and:
\begin{equation}
\frac{A'}{H^2 r} =-\frac{\eta
e\mu^2}{2(1+\beta^2)}\frac{\beta^2-f^2}{{\cal E}_2(f)(1+f^2)}
\label{GravcylindrMax}
\end{equation}
Einstein equations for the metric field (\ref{Einstein2}) get in
the cylindrical case the following form:
\begin{equation}
\left[r \left((\log H)' -\frac{8\pi G}{\eta e}B {\cal E}_2(f)
(\log f)'\right)\right]'=0 \label{EinsteinRad}
\end{equation}

Since we are interested here in cases where the matter fields tend
asymptotically to their vacuum values, the geometry of space will
evidently be conic with a deficit angle given by the usual
relation:
\begin{equation}
\frac{\delta\varphi}{2\pi} = -\lim_{r\rightarrow \infty}
(r(\log H)') \label{DeltaPhi}
\end{equation}
We may obtain more information about the deficit angle by
integrating equation (\ref{EinsteinRad}) once and using the result
at both ends, $r=0$ and $r\rightarrow \infty$:
\begin{equation}
\frac{\delta\varphi}{2\pi} =\frac{8\pi G}{e}|nB(0)| {\cal
E}_2(f(0))\label{BandDeltaphi}
\end{equation}
where we used the boundary conditions (\ref{moreBC}) and
(\ref{BConAatInf}) supplemented by $H(0)=1,\ H'(0)=0$, and the
fact that $-nB(0)$ is always (i.e. for $\eta=\pm 1 $) positive.
The factor $B(0) {\cal E}_2(f(0))$ in the last equation can be
expressed in terms of the dielectric function ${\cal E}_1 (f)$
using (\ref{Bogconstraint}) and assuming a vanishing magnetic
field  as $f\rightarrow\beta$:
\begin{equation}
\frac{\delta\varphi}{2\pi} =8\pi G\mu ^2 |n|\int_0^\beta f{\cal E}_1
(f)df \label{Deltaphi2}
\end{equation}
For  the O(3) sigma model ${\cal E}_1 (f)$ under consideration here, this integral
can be explicitly calculated, or we may rather use the already
given expression of the magnetic field - Eq.(\ref{GravcylindrMax}).
Both ways we get:
\begin{equation}
\frac{\delta\varphi}{2\pi} =\frac{4\pi G v^2 |n|}{1+\beta ^2}
\label{BethaAndDeltaphi}
\end{equation}
Note that these two last equations are valid for the $f(0)=0$
solutions - see final note of previous section. The well-known
usual Higgs result \cite{Linet} is obtained for $\beta=0$, while
holding $v$ fixed. On the other hand, we may take another limit to
obtain the deficit angle for the "anti-Schroers" solutions by
$\beta\rightarrow \infty$, while holding $\mu$ fixed. This gives
$(\delta\varphi/{2\pi})_S =4\pi G \mu^2 |n|$. This result applies
also to the Schroers solutions.

\begin{figure}[!t]
   \begin{center}
   \includegraphics[width=10cm,angle=270]{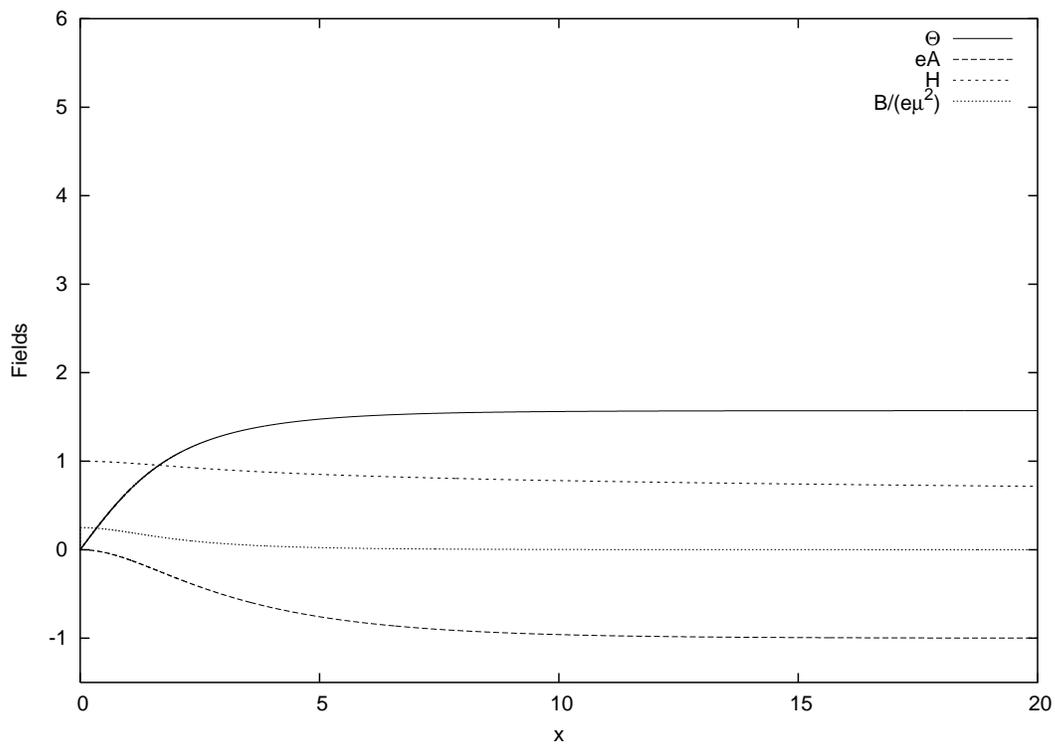} \\
   \caption{ The solution to Eqs.(\ref{Gravcylindrselfduality})-(\ref{EinsteinRad}),
   for ${\cal
E}_2=1$. The parameters used are  $n=-1,\ \beta =1,\ 8\pi
G\mu^2=0.5$. Compare with Figure 1.}
 \label{figure2}
   \end{center}
   \end{figure}

\begin{figure}[!t]
   \begin{center}
   \includegraphics[width=10cm,angle=270]{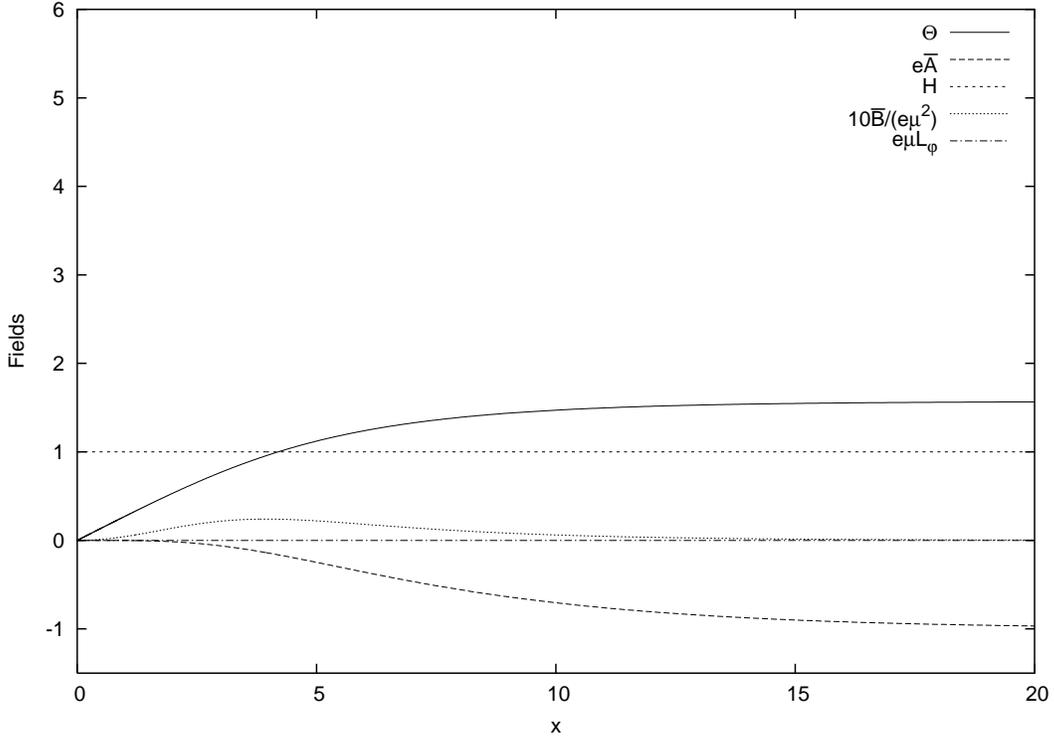} \\
   \caption{The solution to Eqs.(\ref{Lequation})-(\ref{EinsteinCSRad}), with
the boundary condition $f(\infty )= \beta$. The parameters used
are  $n=-1,\ \beta =1,\ \kappa /e\mu=1,\ 8\pi G_3\mu^2=0$, i.e.
gravity is not included. Notice that we magnify the magnetic field
by a factor 10 here and in the subsequent figures. The gauge field
tends to its asymptotic value slower than in the previous case as
is seen from the value of $e\bar{A}$ at $x=20$ which is somewhat
above -1. The curves in this figure represent also the
corresponding solution for the GMH system with the triple well
potential - see remark below Eq. (\ref{relHCSSigma}).}
 \label{figure3}
   \end{center}
   \end{figure}

In Fig.\ref{figure2}, we show the fields ($\Theta, \ A,\ B, \ H$)
for the case $\beta =1$, $8\pi G\mu^2=0.5$; compare with
Fig.\ref{figure1}.

\section{Chern-Simons Type of the Generalized Higgs Model}\label{GCSH Model}
\setcounter{equation}{0}

The possibility of a second field dependent "dielectric function"
${\cal E}_2$, used at the end of section \ref{Flat Space Solutions}, is
much more interesting from another aspect, which is a connection
with $D=3$ Chern-Simons theory. It turns out that the self-dual
solutions for a $D=3$ system of a gauged sigma model coupled to
pure Chern-Simons theory, are related to those of our generalized
Higgs system. In other words, we may replace the Maxwell term with
a Chern-Simons term in the $D=3$ version of the action
(\ref{action1}) to get a
generalized Chern-Simons-Higgs  (GCSH) model. We will
find interesting relations between the self-dual solutions of this
GCSH system and the previous one, which from now on we call
generalized Maxwell-Higgs (GMH) model.

There is of course a physical difference between time-independent
solutions in both models, which is the presence of electric charge
and electric field in the GCSH model. However, due to the linear
relation between the magnetic field and the electric charge
density, a $|\Phi|$-dependent magnetic energy density appears,
which in flat space mimics exactly the Maxwellian energy density
in the GMH system, as will be shown below. This sheds new light on
the well-known self-dual solutions of the Chern-Simons gauged O(3)
sigma model \cite{Ghosh+Ghosh,KimmEtal,ArthurEtAl,Mukh2}. The
self-dual solutions of the Chern-Simons type of the Abelian Higgs
system \cite{HongetAl,Jackiw+Weinberg}, fit also to this framework
by the obvious choice ${\cal E}_1(|\Phi|)=1$. In order to
demonstrate these relations we write the $D=3$ action:
\begin{eqnarray}
S=\int \ud^3x\sqrt{|g|}\left( \frac{1}{2}{\cal E}_1 (|\Phi|)(D_\mu
\Phi)^*(D^\mu\Phi)-U(|\Phi|)+
\frac{\kappa}{4}\frac{\epsilon^{\lambda\mu\nu}}{\sqrt{|g|}}A_\lambda
F_{\mu\nu} +\frac{1}{16\pi G_3}R\right) \label{action2}
\end{eqnarray}
where all the geometrical terms here are 3-dimensional, and
$\kappa$ is a positive parameter (it cannot be a field-dependent
function, as in the GMH case, because it would  break gauge
invariance) with dimensions of 1/length. Actually, the fields here
have different dimensionalities than in $D=4$, but the self-dual
solutions of the GMH system are identical in $D=3$ and $D=4$, so
the comparison becomes trivial.

The field equations of this system are:
\begin{equation}
{\cal E}_1 (|\Phi|) D_\mu D^\mu \Phi + \frac{\Phi^*}{2|\Phi|}
\frac{d{\cal E}_1}{d|\Phi|} D_\mu \Phi D^\mu \Phi +
\frac{\Phi}{|\Phi|} \frac{dU}{d|\Phi|} = 0 \label{KGCS}
\end{equation}
\begin{equation}
\frac{\kappa}{2}\frac{\epsilon^{\lambda\mu\nu}}{\sqrt{|g|}}
F_{\lambda\mu}=j^{\nu} = -\frac {i}{2}e{{\cal E}_1}
(|\Phi|)(\Phi^*(D^{\nu}\Phi) - \Phi (D^{\nu}\Phi)^*) \label{CS}
\end{equation}
\begin{equation}
\frac{1}{8\pi G_3}R_{\mu\nu}+\frac{1}{2}{\cal E}_1 (|\Phi|)\left(
(D_\mu \Phi)^*(D_\nu\Phi)+(D_\nu
\Phi)^*(D_\mu\Phi)\right)-2U(|\Phi|)g_{\mu\nu} = 0\label{EinstCS}
\end{equation}
Spontaneous symmetry breaking occurs if the potential has a circle
of degenerate minima´, just as for the GMH system. The mass of the
Higgs field is given by the same expression as for GMH -
Eq.(\ref{SSBMasses}), but the mass (squared) of the gauge field is
now \footnote{Compare Deser and Yang \cite{DeserYang}.}:
\begin{equation}
m_A^2=\left( \frac{e^2v^2{\cal E}_1(v)}{\kappa} \right)^2
\label{CSSSBMass}
\end{equation}

In order to find static self-dual solutions in flat space, one may
proceed by dropping the curvature terms and look for conditions
for Eq.(\ref{selfduality}) to be satisfied. However, we prefer to
take the more general approach of dealing with the gravitating
system from the beginning. We therefore try to repeat what we did
in the previous sections and require $g_{00}=1$ in a $D=3$ version
of the static metric (\ref{staticMetric}). It turns out that this
cannot be done "naively", but some modifications are required. The
reason is the well-known property of Chern-Simons vortices, which
should carry also electric charge so the Maxwell tensor must
contain in this case also an electric field, and the vortex carries
angular momentum as well. This addition has important consequences
if we treat the gravitational field as dynamical, since the angular
momentum of the Chern-Simons field forces the gravitational field
to be stationary and not simply static. We therefore parametrize
the $D=3$ metric by:

\begin{equation}
ds^2=N^2(x{^k})\left(dt+L_{i}(x{^l})dx{^i} \right)^2- \gamma
_{ij}(x{^k})dx{^i}dx{^j} \label{stationaryMetric}
\end{equation}
and write for the gauge field:
\begin{equation}
A_{\mu}dx^\mu = A_{0}(x{^k}) dt + A_i(x{^k}) dx^i
\label{GaugeField3D}
\end{equation}

In order to get the field equations for stationary solutions, we
introduce also the following notation:
\begin{eqnarray}
\bar{A}_{i} &=& A_{i} - A_{0} L_{i} \nonumber\ \\
\bar{D}_{i} &=& \partial_{i}-ie\bar{A}_{i} \nonumber\ \\
L_{ij}&=&\partial_{i}L_{j}-\partial_{j}L_{i}={\sqrt{|\gamma|}}\epsilon_{ij}\ell
\nonumber\ \\
\bar{F}_{ij}&=&\partial_{i}\bar{A}_{j}-\partial_{j}\bar{A}_{i}
=-{\sqrt{|\gamma|}}\epsilon_{ij}\bar{B}
 \label{notation3D}
\end{eqnarray}
and compute the components of the Ricci tensor:
\begin{eqnarray}
R_{00}=-N\nabla_i \nabla^{i} N -\frac {N^4}{2}\ell^2 \nonumber\ \\
R^{i}_{0}=-\frac{1}{2N\sqrt{|\gamma|}} \epsilon^{ij}\partial_{j}(N^3 \ell) \\
R^{ij}=\frac{1}{N}\nabla^i \nabla^{j} N +\frac
{1}{2}(R(\gamma)-N^2\ell^2)\gamma^{ij} \nonumber\
\label{Ricci3Dstationary}
\end{eqnarray}
where $\nabla_i$ is the covariant derivative with respect to the
two-dimensional metric $\gamma_{ij}$ and $R(\gamma)$ the
corresponding Ricci scalar.

Now we impose $N=1$ and find that the $(00)$ component of Einstein
equations (\ref{EinstCS}) will be satisfied only with the
following condition on the potential:
\begin{equation}
U(|\Phi|) = \frac{e^2}{2}|\Phi|^2 {{\cal E}_1}(|\Phi|)
(A_0)^2-\frac{\ell ^2}{32\pi G_3}
 \label{CSSelfdualPot}
\end{equation}
From the $(ij)$ Einstein equations (or better from the $G^{ij}$
equations), we find the form of the self-duality equation in this
case:
\begin{equation}
 \bar{D}_i\Phi=i\eta {\sqrt{|\gamma|}} \epsilon_{ij} \gamma^{jk} \bar{D}_k\Phi
 \label{CSselfduality}
 \end{equation}
as well as the following expression for the two dimensional Ricci
scalar:
\begin{equation}
R(\gamma)= \ell^2-8\pi G_3 \left({\cal E}_1 (|\Phi|)
\gamma^{ij}(\bar{D}_i \Phi)^*(\bar{D}_j \Phi)+4U(|\Phi|)\right)
 \label{CSRicciSD}
 \end{equation}
Due to the self-duality, the other field equations simplify as
follows. The non-diagonal Einstein equations give:
\begin{equation}
\partial_i\ell =16\pi G_3 \eta e|\Phi| {{\cal E}_1}(|\Phi|) A_0 \partial_i |\Phi|
 \label{CSRi0}
 \end{equation}
 The spatial Chern-Simons equations give:
 \begin{equation}
\partial_i A_0 = -\frac{\eta e}{\kappa} |\Phi| {{\cal E}_1}(|\Phi|) \partial_i
|\Phi|
 \label{CSspacecomp}
 \end{equation}
which is used in the time component of the Chern-Simons equations
to obtain an expression for $\bar{B}$:
\begin{equation}
\bar{B}=(\ell+\frac{e^2}{\kappa}|\Phi|^2{\cal E}_1 (|\Phi|))A_0
 \label{CSBbar}
 \end{equation}
 Finally, the equation for the Higgs field gives another
 functional relation between the potential and the other
 quantities:
\begin{equation}
\eta e|\Phi| {\cal E}_1(|\Phi|) \bar{B} +\frac{dU}{d|\Phi|}-
\frac{e^2}{2} (A_0)^2 \frac{d}{d|\Phi|}({|\Phi|^2\cal E}_1)=0
 \label{HiggsCS}
\end{equation}
Actually it is not an independent equation, since by substitution
of (\ref{CSSelfdualPot}) in (\ref{HiggsCS}), one may get back
(\ref{CSBbar}).

The two differential equations for $\ell(x^k)$ and $A_0(x^k)$ may
be also converted into equations for $\ell(|\Phi|)$ and
$A_0(|\Phi|)$:
\begin{equation}
\frac{d\ell}{d|\Phi|} =16\pi G_3 \eta e|\Phi| {{\cal
E}_1}(|\Phi|)A_0
 \label{liphi},
 \end{equation}
 \begin{equation}
\frac{dA_0}{d|\Phi|} = -\frac{\eta e}{\kappa} |\Phi| {{\cal
E}_1}(|\Phi|)
 \label{A0phi}
 \end{equation}
and we find a simple expression of $\ell$ in terms of $A_0$:
\begin{equation}
\ell =8\pi G_3 \kappa (c- A_0^2)
 \label{lofA0}.
 \end{equation}
where $c$ is an integration constant, which should be non-negative
for solutions with finite angular momentum (where $\ell$ vanishes
asymptotically).

 As in the gravitating GMH model, the right hand side
 of (\ref{CSRicciSD}) is a two-dimensional divergence and we find:
\begin{equation}
R(\gamma)= 16\pi G_3 \kappa \nabla_{i}\left(\frac{1}{\eta
e}\gamma^{ij}A_0 \partial_j \log|\Phi|
+\frac{c\epsilon^{ij}}{\sqrt{|\gamma|}}
 L_{j} \right)
\label{SimplifRicciCSSD}
\end{equation}
If we use a conformally flat metric, (\ref{SimplifRicciCSSD}) can
be rewritten as:
\begin{equation}
\delta^{ij}\partial_i \left( \partial_j\log H - \frac{8\pi G_3
\kappa}{\eta e} A_0 \partial_j \log|\Phi| \right) = 8\pi G_3
\kappa cH^2\ell \label{EinstCSSDconfFlat}
\end{equation}

 Now we consider some special cases that allow self-dual
 solutions. We will then concentrate in rotationally symmetric
 solutions, i.e. gravitating vortices.

 \subsection{${\cal E}_1(\Phi|)=1$}

 This is the Chern-Simons version of the usual Higgs model,
 which we may consider either in flat space or coupled to gravity.
 Integration
 of Eq.(\ref{A0phi}) is trivial and gives:
  \begin{equation}
A_0 = \frac{\eta e}{2\kappa} (v^2 - |\Phi|^2 )
 \label{A0linHflatsol}
 \end{equation}
 where $v$ is an integration constant. In flat space $\ell = 0$ and
 Eq.(\ref{CSSelfdualPot}) gives immediately the
 well-known \cite{HongetAl,Jackiw+Weinberg} sixth order potential:
\begin{equation}
 U(|\Phi|) = \frac{e^4}{8\kappa^2}|\Phi|^2(v^2-|\Phi|^2)^2
 \label{HiggsCSflatSelfdualPot}
\end{equation}
In the case of a gravitating system, we solve for $\ell (|\Phi|)$
and find:
\begin{equation}
\ell (|\Phi|) =\frac{2\pi G_3 e^2}{\kappa}
\left(\sigma-(v^2-|\Phi|^2)^2\right)
 \label{lvsphi}
 \end{equation}
 where $\sigma = 4\kappa^2c/e^2$. The potential is easily found to be:
\begin{equation}
 U(|\Phi|) = \frac{e^4}{8\kappa^2}\left[|\Phi|^2(v^2-|\Phi|^2)^2-
 \pi G_3 \left(\sigma-(v^2-|\Phi|^2)^2 \right ) ^2 \right ]
\label{HiggsCSSelfdualPotgrav}
 \end{equation}
which has an unbounded additional term, thus rendering the whole
system possibly unstable. Notwithstanding the possible
instability, this potential was discussed by several authors
\cite{Valtancoli,London,Clement,Chung2Kim2} and vortex solutions
were also obtained. There is a simple and natural way to cure the
ill-behaved potential, which is generalizing this system to the
O(3) sigma model.

\subsection{${\cal E}_1(\Phi|)=1/(1+|\Phi|^2/\mu^2)^2$}

This is the Chern-Simons gauged O(3) sigma model considered in
flat background by some authors
\cite{Ghosh+Ghosh,KimmEtal,ArthurEtAl,Mukh2,Gladikowski}. Within the
present framework the analysis is straightforward. First we
integrate Eq.(\ref{A0phi}) and find:
\begin{equation}
A_0=\frac{\eta e\mu^2}{2\kappa(1+\beta^2)}
\frac{\beta^2-|\Phi|^2/\mu^2}{1+|\Phi|^2/\mu^2} \label{A0vsPhiCS}
\end{equation}
In flat space $\ell=0$ and Eq.(\ref{CSSelfdualPot}) gives
immediately the following potential:
\begin{equation}
U(|\Phi|)=\frac{e^4\mu^4}{8\kappa^2(1+\beta^2)^2} \frac{|\Phi|^2
\left(\beta^2-|\Phi|^2/\mu^2\right)^2}{(1+|\Phi|^2/\mu^2)^4}
\label{SelfDualPotCSSO3}
\end{equation}
Some authors have already discussed the self-dual solutions in
flat space  for special cases of the potential, which from this
point of view are just special values of $\beta$: $\beta=0$
\cite{Ghosh+Ghosh,ArthurEtAl} and $\beta=1$ \cite{Mukh2}. This
general form of the potential appeared already (in a different
parametrization) in Ref.\cite{ArthurEtAl}, but only the case
$\beta=0$ was discussed there.  Kimm et al. \cite{KimmEtal} have
studied the general case in flat space and found three kinds of
flux tube solutions characterized by the asymptotic value of
$f(r)$, which (for $f(0)=0$)  may be either $0$, $\beta$ or
$\infty$.

If this Chern-Simons gauged O(3) model is coupled to gravity, we
find that the ill-behaved potential from the Higgs system
becomes now bounded from below. First we integrate (\ref{liphi})
either directly or using (\ref{lofA0}) and (\ref{A0vsPhiCS}) to
get:
\begin{equation}
\ell (|\Phi|) =\frac{2\pi G_3 e^2 \mu^4}{\kappa (1+\beta^2)^2}
\left(\lambda-
\frac{(\beta^2-|\Phi|^2/\mu^2)^2}{(1+|\Phi|^2/\mu^2)^2}\right)
 \label{lvsphiSO3}
 \end{equation}
where the role of the integration constant is played by $\lambda$
defined by $\lambda=(1+\beta^2)^2\sigma/\mu^4$. Then we find from
(\ref{CSSelfdualPot}) a new potential with $\lambda$ as an
additional (non-negative) free parameter:
\begin{equation}
U(|\Phi|)= \frac{e^4\mu^4}{8\kappa^2(1+\beta^2)^2}
\left[\frac{|\Phi|^2
\left(\beta^2-|\Phi|^2/\mu^2\right)^2}{(1+|\Phi|^2/\mu^2)^4}-
\frac{\pi G_3 \mu^4}{(1+\beta^2)^2} \left(\lambda-
\frac{(\beta^2-|\Phi|^2/\mu^2)^2}{(1+|\Phi|^2/\mu^2)^2}\right)^2\right]
\label{SelfDualPotCSSO3grav}
\end{equation}
This potential is clearly bounded from below, thus restoring the
stability. Moreover, it has always a local minimum at
$|\Phi|=v=\beta\mu$, which breaks the U(1) local symmetry. The
extremal points at $|\Phi|=0$ and $|\Phi|\rightarrow\infty$, which
are U(1)-symmetric, may be either minima or maxima depending on the
following conditions:
\begin{eqnarray}
\beta^2(1+\beta^2) +4\pi G_3 \mu^2(\beta^4-\lambda) >0
\;\;\;\; \mbox{minimum at} \;\;\;\;\;\;  |\Phi|=0 \nonumber\ \\
1+\beta^2 +4\pi G_3 \mu^2(1-\lambda) >0 \;\;\;\; \mbox{minimum at}
\;\;\;\;\;\; |\Phi|\rightarrow\infty \label{minimaCond}
\end{eqnarray}
The potential of the gravitating Chern-Simons type of the usual
Higgs system, is obtained in the limit $\mu\rightarrow \infty$
(with $\lambda\mu^4$ and $\beta\mu$ kept finite).

Actually a special case of this potential for $\lambda=0$ was
recently obtained by Abou-Zeid and Samtleben \cite{AbouZeid} from
a different direction of three dimensional $N=2$ supergravity
theory. We will see that $\lambda\neq0$ yields interesting
solutions as well.

Now we concentrate in rotationally symmetric
 solutions, i.e. we use the Nielsen-Olesen ansatz including the
 time component of the gauge potential:
\begin{equation}
\Phi=\mu f(r)e^{in\varphi}\  ,\;\;\; A_\mu
dx^\mu=A_{0}(r)dt+A(r)d\varphi \label{cylAnsatzCS}
\end{equation}
 with the additional requirement that all the metric components
 depend on $r$ only.

\begin{figure}[!t]
   \begin{center}
   \includegraphics[width=10cm,angle=270]{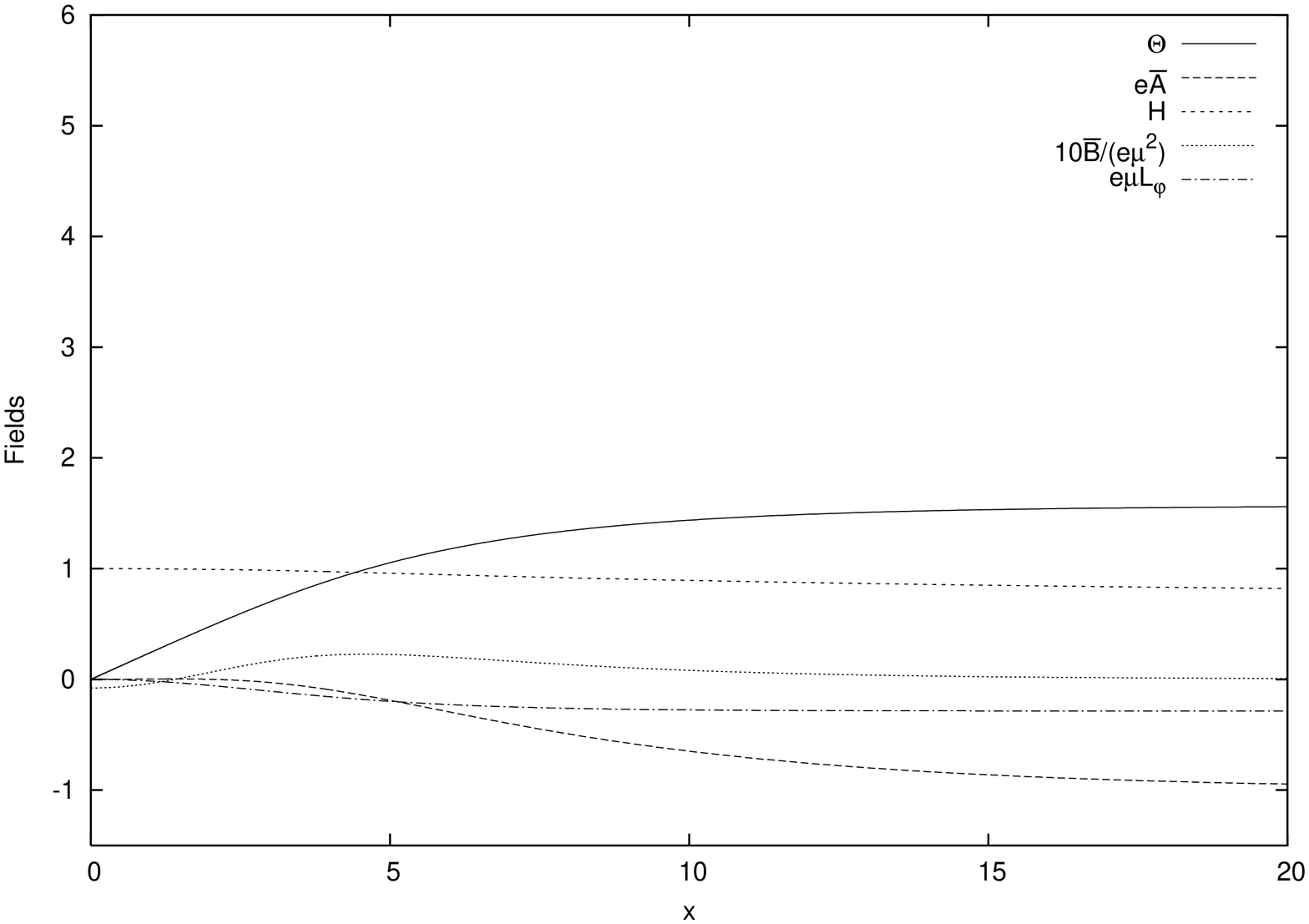} \\
   \caption{The solution to Eqs.(\ref{Lequation})-(\ref{EinsteinCSRad}), with
the boundary condition $f(\infty )=     \beta$. The parameters
used are  $n=-1,\ \beta =1,\ \kappa /e\mu=1,\  8\pi G_3\mu^2=0.5$,
i.e. gravity is included. Compare with Figure 3.}
 \label{figure4}
   \end{center}
   \end{figure}

\begin{figure}[!t]
   \begin{center}
   \includegraphics[width=10cm,angle=270]{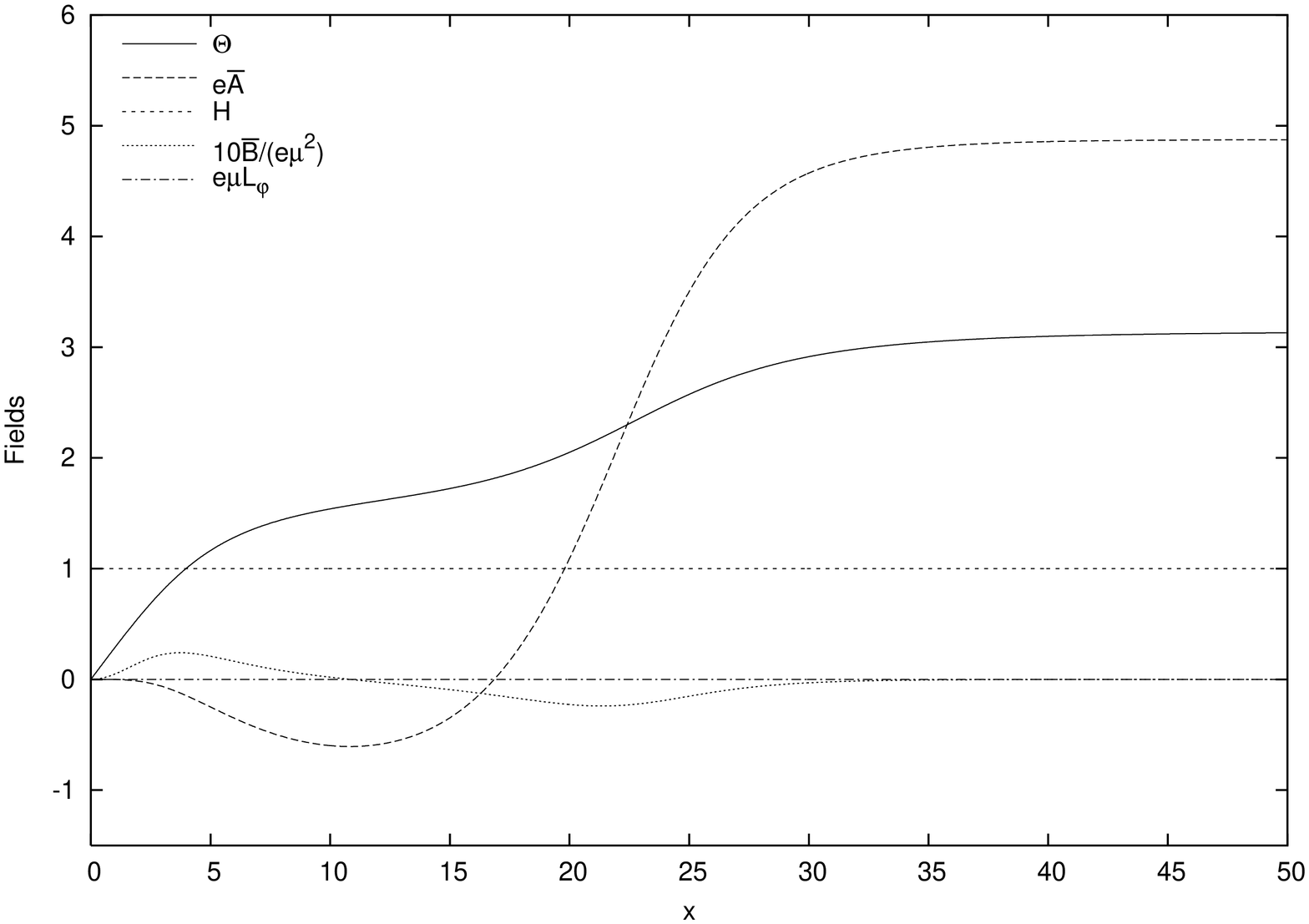} \\
   \caption{The solution to Eqs.(\ref{Lequation})-(\ref{EinsteinCSRad}), with
the boundary condition $f(\infty )= \infty$. The parameters used
are  $n=-1,\ \beta =1,\ \kappa /e\mu=1, \ 8\pi G_3\mu^2=0$, i.e.
gravity is not included. The curves in this figure represent also
the corresponding solution for the GMH system with the triple well
potential - see remark below Eq. (\ref{relHCSSigma}).}
 \label{figure5}
   \end{center}
   \end{figure}

\begin{figure}[!t]
   \begin{center}
   \includegraphics[width=10cm,angle=270]{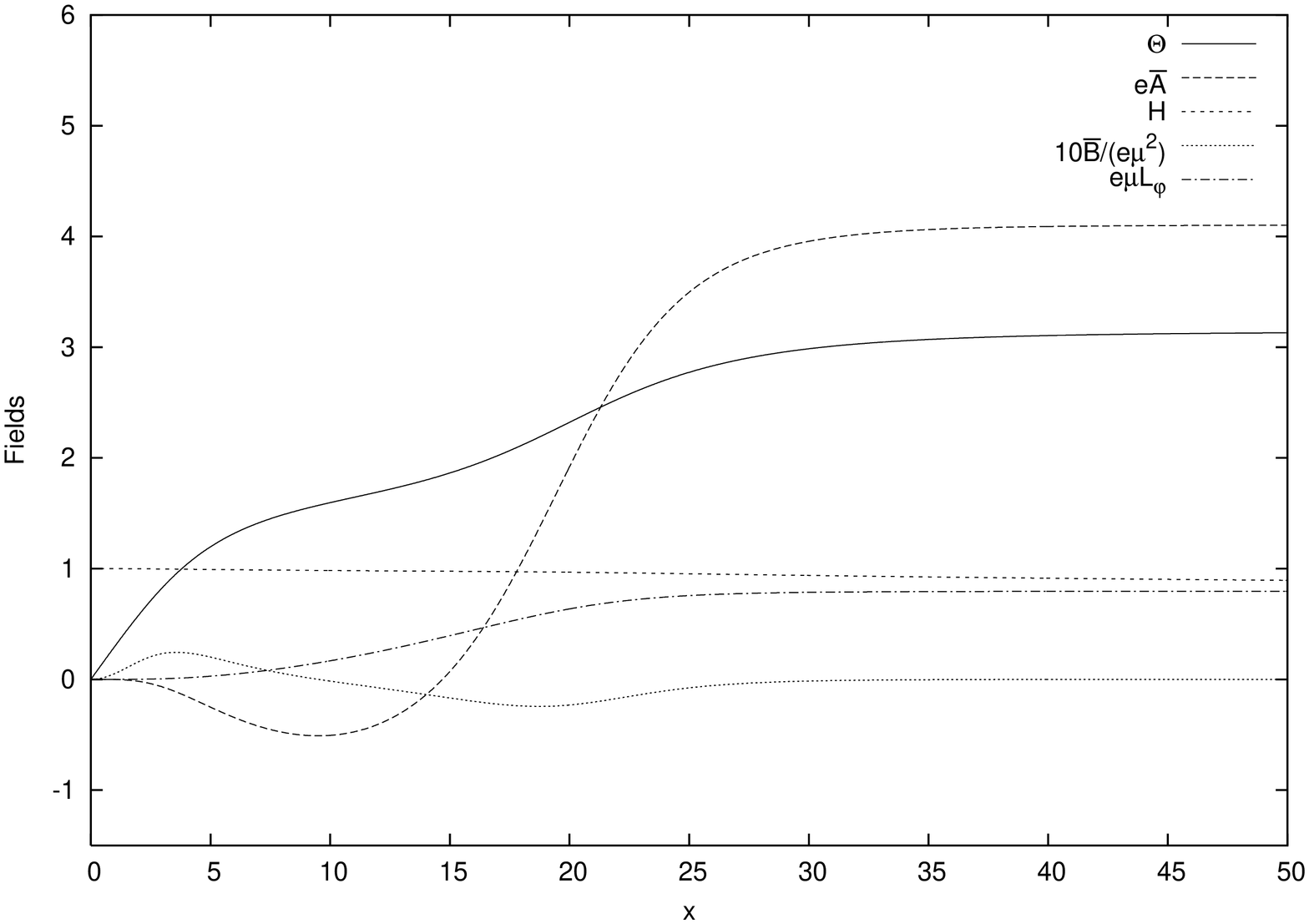} \\
   \caption{The solution to Eqs.(\ref{Lequation})-(\ref{EinsteinCSRad}), with
the boundary condition $f(\infty )=\infty$. The parameters used
are  $n=-1,\ \beta =1,\ \kappa /e\mu=1, \ 8\pi G_3\mu^2=0.0625$,
i.e. gravity is included. Compare with Figure 5.}
 \label{figure6}
   \end{center}
   \end{figure}

\begin{figure}[!t]
   \begin{center}
   \includegraphics[width=10cm,angle=270]{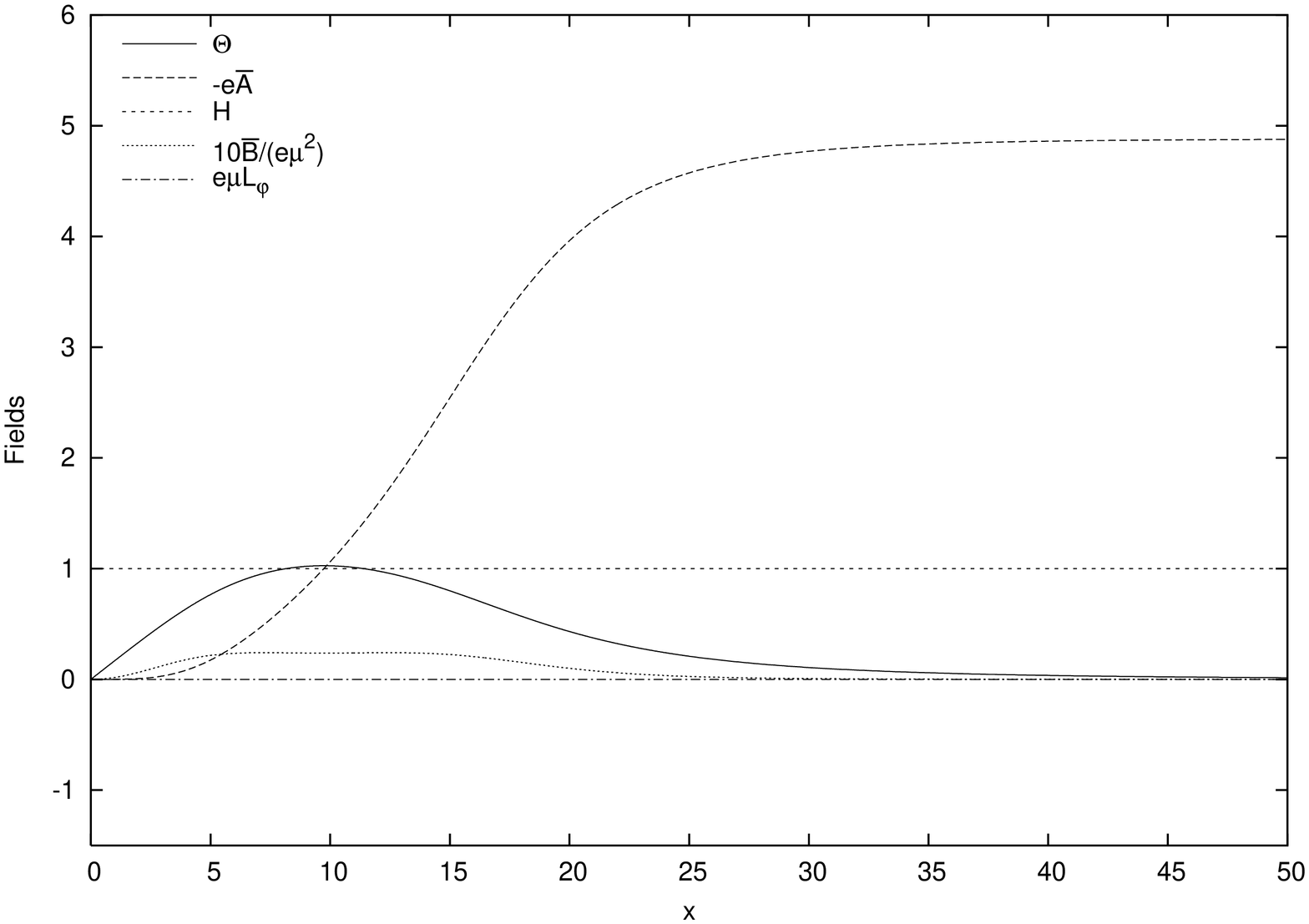} \\
   \caption{The solution to Eqs.(\ref{Lequation})-(\ref{EinsteinCSRad}), with
the boundary condition $f(\infty )=     0$. The parameters used
are  $n=-1,\ \beta =1,\ \kappa /e\mu=1,\  8\pi G_3\mu^2=0$, i.e.
gravity is not included. Notice that we show $-e\bar{A}$. The
curves in this figure represent also the corresponding solution
for the GMH system with the triple well potential - see remark
below Eq. (\ref{relHCSSigma}).}
 \label{figure7}
   \end{center}
   \end{figure}

\begin{figure}[!t]
   \begin{center}
   \includegraphics[width=10cm,angle=270]{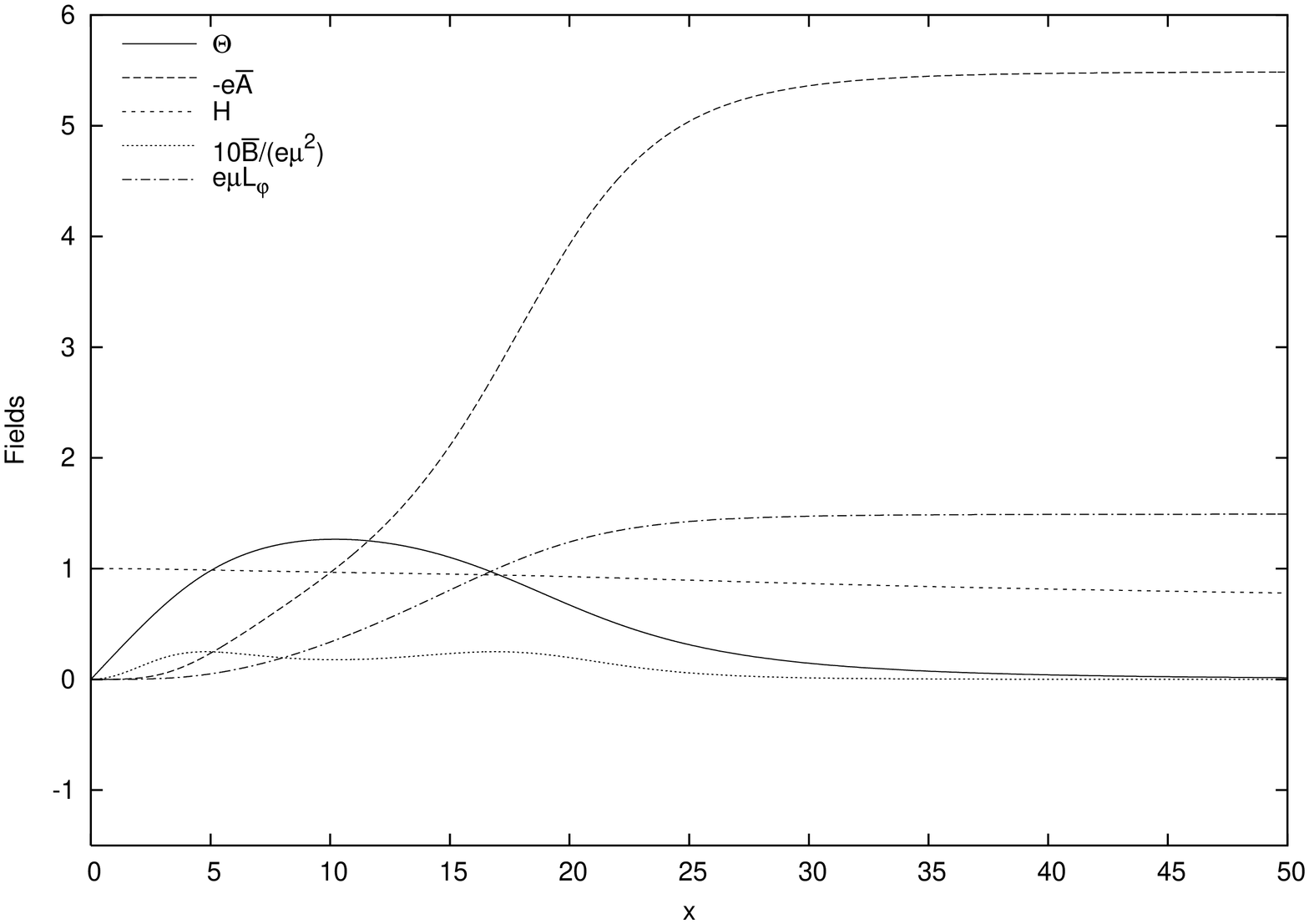} \\
   \caption{The solution to Eqs.(\ref{Lequation})-(\ref{EinsteinCSRad}), with
the boundary condition $f(\infty )=     0$. The parameters used
are  $n=-1,\ \beta =1,\ \kappa /e\mu=1, \ 8\pi G_3\mu^2=0.15$,
i.e. gravity is included.
 Compare with Figure 7}
 \label{figure8}
   \end{center}
   \end{figure}

The field equations for self-dual rotationally symmetric solutions
are quite similar to the ones of the GMH model with two additional
equations: For $A_0$ and for the metric component $L_\varphi (r)$
($L_r$ now vanishes). The dependence of $A_0$ on $f$ is obvious
from Eq.(\ref{A0vsPhiCS}). For $L_\varphi (r)$ we have:
\begin{equation}
\frac{L'_\varphi}{H^2 r}=\ell
 \label{Lequation}
\end{equation}
where the dependence of $\ell$ on $f$ is obtained from
(\ref{lvsphiSO3}). The self-duality equation for the scalar field
is now ($\eta=+1$ as usual):
\begin{equation}
\frac{rf'}{f}=e\bar{A}(r)-n \label{GCSHselfduality}
\end{equation}
where for brevity we denote $\bar{A} = A -A_0 L_\varphi$.
The equation for $\bar{A}(r)$ may be easily obtained from
(\ref{CSBbar}), expressing $\ell$ and $A_0$ in terms of $f$:
\begin{equation}
\frac{\bar{A}'}{H^2 r}
=-\frac{e^3\mu^4}{2\kappa^2(1+\beta^2)}\frac{\beta^2-f^2}{1+f^2}
\left[\frac{2\pi G_3 \mu^2}{(1+\beta^2)^2}
\left(\lambda-\frac{(\beta^2-f^2)^2}{(1+f^2)^2}\right)+
\frac{f^2}{(1+f^2)^2} \right]
 \label{GravcylindrCS}
\end{equation}
Einstein equations for the metric field
(\ref{EinstCSSDconfFlat}) reduce in this case to the following:
\begin{equation}
\left[r \left((\log H)' -\frac{8\pi G_3 \kappa}{e} A_0 (\log
f)'\right)-8\pi G_3 \kappa cL_\varphi \right]'=0
\label{EinsteinCSRad}
\end{equation}
The reflection symmetry observed in the GMH system exists in the
GCSH system as well, provided $\lambda$ is also rescaled according
to its $\beta$ dependence. We may therefore limit ourselves to
solutions with $f(0)=0$ as before. Clearly there are three kinds
of solutions classified by the three possible values which
$f(\infty)$ may take, according to the different minima of the
potential. However, these three values $U(0)$, $U(v)$ and
$U(\infty)$ are generally different from each other, and each of
them vanishes for a different value of $\lambda$. Since vanishing
value of the potential minimum is an essential property of our
localized solutions, we encounter here a situation which differs
from that in flat space \cite{KimmEtal}, where all three possible
boundary conditions are realized for the same potential. In the
gravitating case, each kind is realized for a different value of
$\lambda$. The flux tube solutions with quantized flux and the
usual boundary condition $f(\infty)=\beta$ exist for $\lambda = 0$
only. Representative solutions are depicted in
Figs.\ref{figure3}-\ref{figure4}. The two other solution types
with $f(\infty)=0,\infty$ exist for $\lambda=\beta^4, 1$,
respectively and are depicted in Figs.\ref{figure5}-\ref{figure8}.
We may therefore  summarize the situation by the following
relations:
\begin{eqnarray}
\lambda = \beta^4 \;\;\;\; &\Rightarrow& \;\;\;\;\;\;
\lim_{r\rightarrow \infty} f(r)=0 \nonumber\\
\lambda = 0 \;\;\;\; &\Rightarrow& \;\;\;\;\;\; \lim_{r\rightarrow \infty} f(r)=\beta \\
\lambda = 1 \;\;\;\; &\Rightarrow& \;\;\;\;\;\; \lim_{r\rightarrow
\infty} f(r)=\infty \nonumber \label{BCatInflambda}
\end{eqnarray}
A special case is $\lambda = \beta = 1$ which allows both boundary
conditions ($f(\infty)=0,\infty$) for the same potential.

Note that all three kinds of boundary conditions ensure also a
vanishing asymptotic value of the angular momentum density $\ell$,
which is a necessary condition for a finite angular momentum. The
gauge potential $A_0$ need not always vanish asymptotically. It
does only in the $f(\infty)=\beta$ case.

For all these three kinds of boundary conditions, the matter fields
tend asymptotically to their vacuum values. Thus asymptotically,
the geometry of space-time will be rotating conic metric.  The
angular deficit may be easily obtained from (\ref{EinsteinCSRad})
to be:
\begin{equation}
\frac{\delta\varphi}{2\pi} =-\frac{8\pi G_3 \kappa}{e} \left[n
A_0(0)+A_0(\infty)(eA(\infty)-n)\right] \label{AandDeltaphiCS}
\end{equation}
For vanishing $c$ (or $A_0(\infty)$ or $\lambda$, i.e. quantized
flux) we find:
\begin{equation}
\frac{\delta\varphi}{2\pi} =-\frac{8\pi G_3 \kappa}{e} n A_0(0)
=8\pi G_3 \mu ^2 |n|\int_0^\beta f{\cal E}_1 (f)df
\label{DeltaphiCSlam0}
\end{equation}
that is the same as the result (\ref{Deltaphi2}) for the GMH
model. For the O(3) case under consideration here, we find the same
expression as we had in the case of
O(3)-GMH:
\begin{eqnarray}
\frac{\delta\varphi}{2\pi} &=&\frac{4\pi G_3 v^2 |n|}{1+\beta ^2}
 \;\;\;\;,\;\;\;\;
\lim_{r\rightarrow \infty} f(r)=\beta \label{deltaphiCSq}
\end{eqnarray}
The angular deficit for the other solutions without flux
quantization (i.e. non-vanishing $c$ or $\lambda$) can be easily
calculated in the O(3) model from (\ref{AandDeltaphiCS}):
\begin{eqnarray}
\frac{\delta\varphi}{2\pi} &=& \frac{4\pi G_3 \mu^2 \beta^2
e|A(\infty)|}{1+\beta ^2}  \;\;\;\;,\;\;\;\;
\lim_{r\rightarrow \infty} f(r)=0 \\
\frac{\delta\varphi}{2\pi} &=& 4\pi G_3 \mu^2 \left|
\frac{eA(\infty)}{1+\beta ^2}-n \right| \;\;\;\;,\;\;
\lim_{r\rightarrow \infty} f(r)=\infty \label{deltaphiCSnq}
\end{eqnarray}
Note the dependence on the unquantized magnetic flux represented
here by $A(\infty)$. Thus we have in these cases a continuum of
values for the angular deficit for a given potential and $n$ and
$f(\infty)$ values.

\vspace{8mm}
We end this section by computing the angular momentum
$J$ of the rotationally symmetric solutions, which we get by
integrating $T^0_\varphi$ over all two-space. The angular momentum
also determines the asymptotic value of the metric component
$L_\varphi (r)$ such that:
\begin{equation}
(1-\frac{\delta\varphi}{2\pi}) L_\varphi(\infty) = -4GJ
\label{LJRelation}
\end{equation}
This relation is most easily obtained in a Gaussian normal coordinate system, in
which the line element is written as:
$ds^2=-d\rho^2+h_{\alpha \beta}dx^{\alpha}dx^{\beta}$
($x^{\alpha} = t,\varphi$). Since $h_{\alpha \beta}$
depends only on $\rho$, the
components of the Ricci tensor have the following simple form in terms of the
extrinsic curvature $K^\alpha _\beta$:
\begin{equation}
R^\alpha _\beta = \frac{1}{\sqrt{|h|}}\frac{\partial}{\partial
\rho} (\sqrt{|h|} K^\alpha _\beta) ,\;\;\;\ K^\alpha
_\beta=-\frac{1}{2} h^{{\alpha \gamma}}\frac{\partial h_{{\gamma
\beta}}} {\partial \rho} \label{RicciNormalCo}
\end{equation}
It turns out that we can compute the angular momentum directly
without even the self-duality assumption, due to the simple
identity which holds for solutions of the flux tube form
(\ref{cylAnsatzCS}):
\begin{equation}
eT^0_\varphi = (n-eA(r))j^0
\label{TophiRelation}
\end{equation}
Now we use the time component of the Chern-Simons equation (\ref{CS}), to
write $j^0$ in terms of $B$ and get:
\begin{equation}
J=\frac{2\pi \kappa}{e}\int_0^\infty (n-eA(r))A'dr=
\frac{\pi \kappa}{e}(2n-eA(\infty))A(\infty)
\label{AngMom}
\end{equation}
Solutions which break the U(1) symmetry asymptotically have quantized angular
momentum of:
\begin{equation}
J=\frac{\pi \kappa}{e^2}n^2
\label{QuantAngMom}
\end{equation}
while otherwise, there is no flux quantization and we cannot say
about the angular momentum more than the right hand side of
(\ref{AngMom}). Note that the angular momentum in
(\ref{QuantAngMom}) is a very general result independent of all
details of the theory, except the existence of symmetry breaking
vacuum in the potential. We note further that all solutions have
the same sign of angular momentum, which is a manifestation of
parity violation.

\section{Correspondence between Maxwell and Chern-Simons Type
 of the Generalized Higgs Model}
 \label{correspondence}
\setcounter{equation}{0}

We note that in flat space, the electric energy density of the
Chern-Simons field plays an equivalent role to that of the
magnetic energy density in the GMH model. Moreover, due to
(\ref{CSBbar}), there is a relation between the magnetic and
electric terms, thus enabling us to eliminate the scalar (electric)
potential from the equations and to give them a form identical to
those of the GMH model.

 Comparison of the Bogomolny constraints in both cases,
 equations (\ref{Bogconstraint}) and (\ref{A0phi}), gives a relation between the
 magnetic field in the GMH model and the time component of the gauge
 potential in the GCSH model:
\begin{equation}
{\cal E}_2 B=\kappa A_0  \label{relA0B}
\end{equation}
Further comparison of the magnetic field appearing in this
equation with the one in (\ref{CSBbar}) (with $\ell = 0$) or the
potentials ((\ref{equB}) and (\ref{CSSelfdualPot}) (with $\ell =
0$)), yields the following characteristic relation for the
generalized Maxwell-Higgs model:
\begin{equation}
\frac{\kappa^2}{{\cal E}_2(|\Phi|)}=e^2|\Phi|^2{\cal E}_1(|\Phi|)
\label{relHCSSigma}
\end{equation}
Thus, all the flat space self-dual solutions to the $D=3$ system,
studied in section 5, are also self-dual solutions to the $D=4$
GMH model of section \ref{Flat Space Solutions}, provided we use
${\cal E}_2$ consistent with (\ref{relHCSSigma}). Consequently,
the curves in figures \ref{figure3},\ref{figure5},\ref{figure7}
represent also the corresponding fields of the GMH model with the
triple well potential (\ref{SigmaTripleWell}) and the appropriate
dielectric function mentioned together with that potential.

However, the presence of an electric field in the $D=3$
Chern-Simons theory, in addition to the magnetic field, is
responsible for the existence of angular momentum, which calls for
corrections to this simple correspondence when gravity is
considered dynamical. It is straightforward (although tedious) to
show that there does not exist spinning gravitating self-dual flux
tubes in the GMH model. Therefore, we have only static versus
stationary correspondence, and all we can hope for is equivalence
of the two-dimensional spatial metrics in addition to the scalar
and vector correspondence. This turns out to be the case, provided
we also impose $c=\sigma=\lambda =0$.

Equation (\ref{relA0B}) is evidently still valid in the
gravitating case and should be accompanied by $B=\bar{B}$ where
$B$ is the GMH magnetic field defined by (\ref{Maxwell2Form}) and
$\bar{B}$ - in (\ref{notation3D}). Thus we get the relation
instead of (\ref{relHCSSigma}):
\begin{equation}
\frac{\kappa^2}{{\cal E}_2(|\Phi|)}=e^2|\Phi|^2{\cal
E}_1(|\Phi|)+\kappa \ell \label{relHCSSigmagrav}
\end{equation}
Note that now the potentials are not equal in both cases, but there
is an $\ell$-dependent difference:
\begin{equation}
U_{M} = U_{CS}-\frac{\ell ^2}{32\pi G_3} \label{PotentialDiff}
\end{equation}

As for the metric tensor, it is easy to see that both expressions
for the Ricci scalar, equations  (\ref{SimplifRicciNLHSD}) and
(\ref{SimplifRicciCSSD}), become identical due to equation
(\ref{relA0B}), thus resulting in equal 2-metrics in the two  cases.

Actually, those $\ell$-dependent modifications should be expressed
in terms of functions that appear in the Lagrangian,  and should be
considered as functions of $|\Phi|$. We will not do this
explicitly, since it is straightforward and does not add any more
insight into the physical picture.

\section{Topological Charge}
\setcounter{equation}{0}
\label{Topological Charge}

The quantization of magnetic flux is intimately connected with the
fact that the minimum of the potential breaks the local U(1)
symmetry. Consequently, the magnetic flux number is the index, or
winding number, of the map defined by the scalar fields from
infinite distance from the flux tube to the vacuum manifold.

If however, the target space is compact, there exists a further
possibility of homotopy classification of the maps defined by the
fields from "real" space to target space. We can use, as a ground
of the present discussion, the well-known notions from sigma models
defined in three spacetime dimensions \cite{Rajaraman}. The GCSH
model is naturally three dimensional, but the GMH model should be
thought of here as three dimensional as well. In this case, the scalar
fields map the two dimensional physical space into target space.
We have therefore an identically conserved topological current
defined by:
\begin{equation}
K^{\lambda}= -\frac{\epsilon^{\lambda\mu\nu}}{2\Omega_T
\sqrt{|g|}} \left(i{\cal E}_1(|\Phi|)(D_\mu \Phi)^*(D_\nu \Phi)+e
{\cal F}(|\Phi|) F_{\mu\nu}\right) \label{TopCurrent}
\end{equation}
where $\Omega_T$ is the target space volume, and ${\cal F}(|\Phi|)$
is defined by:
\begin{equation}
\frac{d{\cal F}}{d|\Phi|} = |\Phi| {\cal E}_1(|\Phi|)
\label{calF}
\end{equation}
This is a gauge invariant generalization of the standard
non-gauged sigma model current, whose time component integral over
all space is the winding number, or index, of the map defined by the
sigma model fields. Note that ${\cal F}(|\Phi|)$ is connected to
${\cal A}(|\Phi|)$ (defined in Eq.(\ref{superdielectric2nd})) by
$e{\cal F}(|\Phi|) = -|\Phi|d{\cal A}/{d|\Phi|}$, up to a possible
additive constant. By the field equations of the GCSH model, it is
also related to $A_0$ by $\kappa A_0 (|\Phi|)=-\eta e{\cal
F}(|\Phi|)$, up to an additive constant. Unlike previous authors
\cite{Mukh1,KimmEtal,ArthurEtAl,Mukh2}, we fix the additive
constant of ${\cal F}$ such that ${\cal F}$ vanishes as
$|\Phi|\rightarrow \infty$, irrespective of the boundary conditions
imposed on the solutions. This way, we can treat uniformly all
sigma models differing only by the potential term. For the special
O(3) "dielectric function", given in (\ref{so3conformalfactor}), we
may write the topological charge (and current) in terms of ${\cal
E}_1(|\Phi|)$ only, since ${\cal F}(|\Phi|)=-\mu^2 \sqrt{{\cal
E}_1(|\Phi|)}/2$. In this case, we have also $\Omega_T=\pi\mu^2$
and the apparent dependence on the scale $\mu$ disappears.

The fact that this gauge invariant current has still a topological
meaning is clearly seen by the fact that it is actually the
non-gauged sigma model current with an addition, which is a
divergence of an anti-symmetric tensor:
\begin{equation}
K^{\lambda}= -\frac{\epsilon^{\lambda\mu\nu}}{2\Omega_T
\sqrt{|g|}} \left( i{\cal E}_1(|\Phi|)(\partial_\mu
\Phi)^*(\partial_\nu \Phi)+2e\nabla_\mu( {\cal F}(|\Phi|) A_{\nu})
\right) \label{TopCurrentNonC}
\end{equation}
Indeed, the first term of (\ref{TopCurrentNonC}) can also be
written as a divergence of an anti-symmetric tensor, so we have the
alternative form:
\begin{equation}
K^{\nu}= \nabla_\mu K^{\mu\nu},\;\;\;\; K^{\mu\nu}=  \frac{{\cal
F}(|\Phi|)\epsilon^{\mu\nu\lambda}}{2\Omega_T \sqrt{|g|}}
\left(\frac{i}{|\Phi|^2}(\Phi^*\partial_\lambda
\Phi-\Phi\partial_\lambda \Phi^*)+2e A_{\lambda} \right)
\label{TopCurrentasDiv}
\end{equation}

Integration of the time component of the topological current
provides us with the topological charge $T$, which characterizes
the solutions from the homotopy point of view. For our stationary
solutions, we write the two equivalent expressions:
\begin{eqnarray}
T= -\frac{1}{2\Omega_T}\int \ud^2x \left(i\epsilon^{ij} {\cal
E}_1(|\Phi|)(D_i \Phi)^*(D_j \Phi)-2e\sqrt{|\gamma|}B{\cal
F}(|\Phi|) \right)  \nonumber\ \\ =-\frac{i}{2\Omega_T}\int
\ud^2x \epsilon^{ij} {\cal E}_1(|\Phi|)(\partial_i
\Phi)^*(\partial_j \Phi) - \frac{e}{\Omega_T} {\cal
F}(w)\oint_{r\rightarrow\infty} A_i dx^i
 \label{TopCharge}
\end{eqnarray}
where $w$ is the asymptotic value of $|\Phi|$. We stress the
difference between $w$, which corresponds to a property of a
solution, and $v$ which is a parameter in the potential. $w$ need
not be equal to $v$ (although it may be), as happens for solutions
of the GCSH model discussed in section \ref{GCSH Model}, or solutions
to the GMH model with the potential (\ref{SigmaTripleWell}), which
may have all three asymptotic values $w=0,v,\infty$.

 Note that $T$ cannot be expressed as a surface integral (actually
 line integral in $D=2+1$) since the winding number, i.e. the
(non-gauged) sigma model contribution of $K^{i0}$, does not
satisfy Stokes theorem. The second term is however a boundary
term, and we note that it vanishes for solutions for  which
$|\Phi|\rightarrow \infty$ as $r\rightarrow \infty$. In the O(3)
model,  these are solutions which do not break the U(1) symmetry
asymptotically. Another possibility is solutions which do break
the U(1) symmetry asymptotically, and thus do not cover all target
space. This kind of solutions have non-integer winding number, but
the topological charge has also a magnetic flux contribution,
which may compensate for this fact.

This is exactly what happens for the self-dual cylindrically
symmetric solutions of the O(3) GMH model, where we find that in
all cases $T=n$. If the solution tends to a symmetry breaking
vacuum, the boundary term of (\ref{TopCharge}) does not vanish and
evaluates to $n/(1+\omega^2)$, where we define $\omega=w/\mu$. The
winding number, which is the first term, is easily found to be
$n\omega^2/(1+\omega^2)$. They both add up to give $T=n$. In the
case $\omega\rightarrow\infty$, where the solution does not break
asymptotically the U(1) symmetry, the flux contribution vanishes
but the first is an integer and we get $T=n$ again. The third
possibility, $\omega=0$ is however an exception: In this case, the
winding number vanishes but we have a flux contribution to get
$T=eA(\infty)$.

Another property, which we can note here, is the relation with the
Euler number of the 2-surface generated by the self-dual GMH
solutions. The Euler number is given by:
\begin{equation}
\chi=-\frac{1}{4\pi}\int d^2x \sqrt{|\gamma|} R(\gamma)
 \label{Euler}
\end{equation}
while for (say) self-dual solutions ($\eta =1$), we may write the
topological charge as:
\begin{eqnarray}
T= -\frac{1}{2\Omega_T}\int \ud^2x \sqrt{|\gamma|}\left({\cal
E}_1(|\Phi|)\gamma^{ij}(D_i \Phi)^*(D_j \Phi)+
2{\cal E}_2(|\Phi|)B^2 \right) - \frac{e}{\Omega_T} {\cal
F}(v)\oint_{r\rightarrow\infty} A_i dx^i
 \label{TopChargeSD}
\end{eqnarray}
By the field equations (\ref{equB}) and (\ref{RicciNLHSD}), we see
that the first term is just proportional to the Euler number, so we
have a simple relation between the topological charge, the Euler
number and the magnetic flux, which we denote here by $\Psi$:
\begin{equation}
 T+\frac{\chi}{4G\Omega_T}-\frac{e{\cal
F}(v)\Psi}{\Omega_T}=0
 \label{TopEulerFlux}
\end{equation}
For the O(3) system, (\ref{TopEulerFlux}) yields the following relation:
\begin{equation}
2\pi T+\frac{\chi}{2G\mu^2}+\frac{e\Psi}{1+\beta^2}=0
 \label{TopEulerFluxO3}
\end{equation}
For asymptotically conic space, the Euler number is related to
the deficit angle by $\chi = \delta\varphi /2\pi$, so
Eq.(\ref{TopEulerFlux}) is easily verified by using
(\ref{BethaAndDeltaphi}).

Next we turn to the GCSH model, where things are quite similar. The
topological current and charge are still given by
Eqs.(\ref{TopCurrent})-(\ref{TopCharge}). However, since the
self-duality condition is now modified by replacing $A_i$ by
$\bar{A}_{i}$, it is useful to write the topological charge also
in a way that is ready for direct use of the modified
self-duality. We therefore write the topological charge density as:
\begin{equation}
K^{0}= -\frac{\epsilon^{ij}}{2\Omega_T \sqrt{|g|}} \left(i{\cal
E}_1(|\Phi|)(\bar{D}_i \Phi)^* (\bar{D}_j \Phi)+
2e\partial_i({\cal F}(|\Phi|) A_0 L_{j})\right) + \frac{e{\cal
F}(|\Phi|)\bar{B}} {\Omega_T} \label{TopDensity}
\end{equation}
Note however that the surface term vanishes only for quantized
flux solutions ($c=\lambda=0$, $A_0 (w)=0$). If we proceed along
similar lines as for the GMH case, we easily find the same relation
between the topological charge, Euler number and the magnetic flux,
i.e. Eq.(\ref{TopEulerFlux}). It simplifies again for the O(3)
model to (\ref{TopEulerFluxO3}).

Finally, we use this relation in order to calculate the
topological charge for the solutions with quantized as well as
non-quantized flux, which is obtained directly from
Eq.(\ref{TopCharge}):
\begin{eqnarray}
T &=& \left\{\begin{array}{ll} eA(\infty) \;\;  &,\;\;\;\; w=0 \\
 n \;\; &,\;\;\;\;  w=v \\
 n \;\; &,\;\;\;\;  w\rightarrow\infty \end{array}\right. \label{TopChCS}
\end{eqnarray}
A nice check of our results is that we can easily reproduce by
substitution in (\ref{TopEulerFluxO3}) the expressions
(\ref{deltaphiCSq})-(\ref{deltaphiCSnq}) for the deficit angles.

\section{Outlook}
\setcounter{equation}{0}

We have shown that the inclusion of dielectric functions gives the
possibility of dramatic changes in the long range behavior of
scalar, gauge and metric fields around a cosmic string or vortex.
This was obtained even for very simple and symmetric models, like
O(3) with the potential (\ref{SigmaTripleWell}). We believe that
the astrophysical and cosmological consequences deserve further
studies.

Another continuation and generalization of our work is to consider
theories with a more extended field content. One natural
possibility is to gauge a U(1)$\times$U(1) subgroup of the global
symmetry group of a CP(2) non-linear sigma model noticing that
CP(2) can be parametrized by two complex coordinates. One can
therefore expect that in analogy with the way the O(3) model (with
$\mbox{\textrm{S}}^2\sim$ CP(1) as target space) generalizes the
ordinary cosmic string, the gauged CP(2) model generalizes the
superconducting cosmic string \cite{Witten,Carter}.

\vspace{8mm}


\newpage
   
\end{document}